\documentclass[aps,twocolumn,superscriptaddress,longbibliography,floatfix,nofootinbib]{revtex4-2}

\usepackage{nameref}
\usepackage[colorlinks=true,urlcolor=blue,citecolor=blue,allcolors=blue]{hyperref}

\usepackage{color}
\usepackage{amsmath}
\usepackage{esint}
\usepackage{amssymb}
\usepackage{physics}
\usepackage{braket}
\usepackage{bm}
\usepackage{graphicx}
\usepackage{tikz}
\usetikzlibrary{calc}
\usepackage{soul}
\usepackage{xcolor}
\usepackage{array}
\usepackage{slashed}
\usepackage{upgreek}
\usepackage{ulem}
\usepackage{MnSymbol} 
\usepackage[capitalize,noabbrev]{cleveref} 
\usepackage{float}
\usepackage{glossaries}
\glsdisablehyper % Disables acronym hyperlinks (they only work with a glossary, otherwise they just point to the beginning of the paper)
\crefname{equation}{}{} % \cref{eq:...} will only write "(1)" not "Equation (1)"

\renewcommand{\O}{\mathcal{O}}

\renewcommand{\H}{\mathcal{H}}

\renewcommand{\P}{\mathcal{P}}

\newcommand{\KL}[1]{\textcolor{red}{[#1]}}

\newacronym{LGT}{LGT}{lattice gauge theory}
\newacronym{VQE}{VQE}{\textit{variational quantum eigensolver}}
\newacronym{NISQ}{NISQ}{\textit{noisy-intermediate scale quantum}}
\newacronym{HEA}{HEA}{\textit{hardware-efficient ansatz}}

\begin{document}
\title{Enhancing Variational Quantum Eigensolvers for SU(2) Lattice Gauge Theory via Systematic State Preparation}

\author{Klaus Liegener}
\affiliation {Walther-Meissner-Institute, Walther-Meissner-Strasse 8, 85748 Garching, Germany}

\author{Dominik Mattern}
\affiliation{Technical University of Munich, TUM School of Natural Sciences, Department of Physics, 85748 Garching, Germany}

\author{Alexander Korobov}
\affiliation{Technical University of Munich, TUM School of CIT, Department of Computer Science, 85748 Garching, Germany}

\author{Lisa Kr\"uger}
\affiliation {Walther-Meissner-Institute, Walther-Meissner-Strasse 8, 85748 Garching, Germany}
\affiliation{Technical University of Munich, TUM School of Natural Sciences, Department of Physics, 85748 Garching, Germany}

\author{Manuel Geiger}
\affiliation{Technical University of Munich, TUM School of CIT, Department of Computer Science, 85748 Garching, Germany}

\author{Malay Singh}
\affiliation {Walther-Meissner-Institute, Walther-Meissner-Strasse 8, 85748 Garching, Germany}

\author{Longxiang Huang}
\affiliation {Walther-Meissner-Institute, Walther-Meissner-Strasse 8, 85748 Garching, Germany}

\author{Christian Schneider}
\affiliation {Walther-Meissner-Institute, Walther-Meissner-Strasse 8, 85748 Garching, Germany}
\affiliation{Technical University of Munich, TUM School of Natural Sciences, Department of Physics, 85748 Garching, Germany}

\author{Federico Roy}
\affiliation {Walther-Meissner-Institute, Walther-Meissner-Strasse 8, 85748 Garching, Germany}
\affiliation{Institute for Quantum Computing Analytics (PGI-12), Forschungszentrum J\"ulich, 52425 J\"ulich, Germany}

\author{Stefan Filipp}
\affiliation {Walther-Meissner-Institute, Walther-Meissner-Strasse 8, 85748 Garching, Germany}
\affiliation{Technical University of Munich, TUM School of Natural Sciences, Department of Physics, 85748 Garching, Germany}
\affiliation{Munich Center for Quantum Science and Technology (MCQST), 80799 Munich, Germany}

\date{\today}

\begin{abstract}
Computing the vacuum and energy spectrum in non-Abelian, interacting lattice gauge theories remains an open challenge, in part because approximating the continuum limit requires large lattices and huge Hilbert spaces. To address this difficulty with near-term quantum computing devices, we adapt the variational quantum eigensolver to non-Abelian gauge theories. We outline scaling advantages when using a spin-network basis to simulate the gauge-invariant Hilbert space and develop a systematic state preparation ansatz that creates gauge-invariant excitations while alleviating the barren plateau problem. We illustrate our method in the context of SU(2) Yang-Mills theory by testing it on a minimal toy model consisting of a single vertex in 3+1 dimensions. In this toy model, simulations allow us to investigate the impact of noise expected in current quantum devices.
\end{abstract}

\maketitle

%\tableofcontents

\section{Introduction}
One of the most prominent open questions in modern quantum field theories is the mass gap problem \cite{JaffeWitten2000-YangMillsMassGap}. In essence, it asks whether the energy gap between the vacuum (i.e., ground state) and the first excited state remains finite in the continuum limit for interacting theories in $d=3$ (spatial) dimensions. Although progress has been made for some toy models \cite{Balaban1984I,PhysRevLett.123.042002,FontanaTrombettoni2025}, in general, numerically approximating the ground state and its energy remains computationally demanding, especially on large lattices. Simulations of {\it lattice gauge theories} (LGT) \cite{BrowerChrist2018LatticeQCDExascale,PhysRevLett.112.201601} have steadily advanced in recent decades through the development of tools such as tensor network methods \cite{SilviRicoCalarcoMontangero2014,Verstraete2008}. However, %recent investigations point towards the presence of magic in non-Abelian LGT \cite{EspositoCepollaroCappielloHamma2025}. It indicates that 
as these methods break down for highly entangled states in dimensions $d\geq 2$, new computational approaches are required.

\begin{figure}
    \centering
    \includegraphics[width=1.0\linewidth]{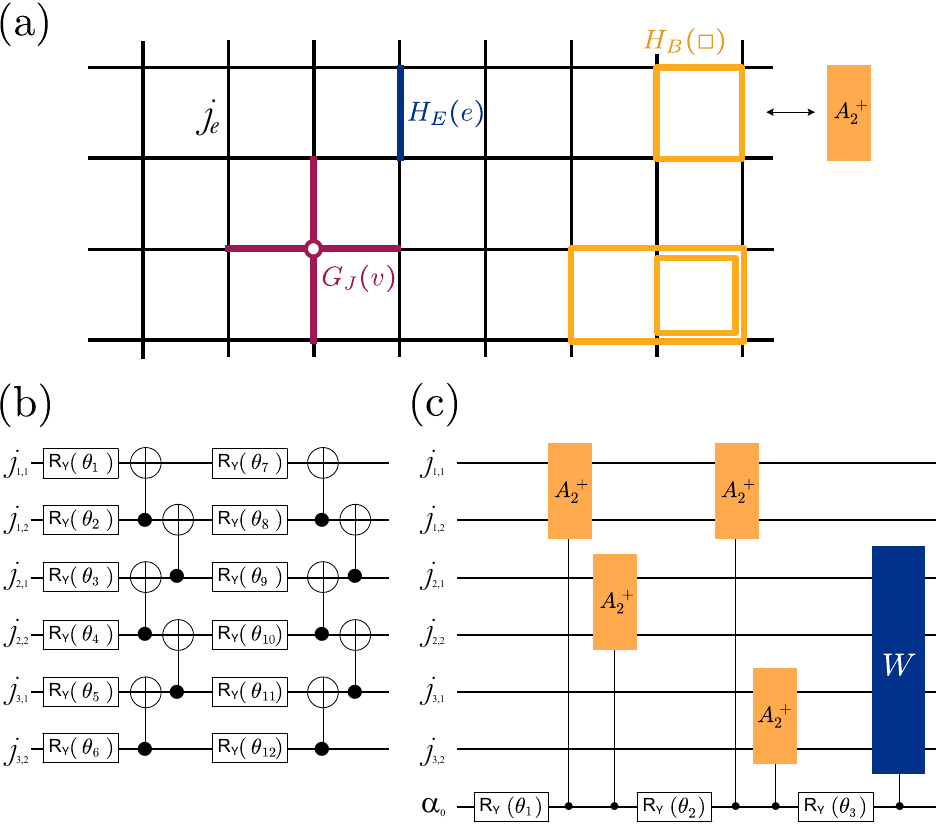}
    \caption{(a) Schematic representation of an LGT system. A Gauss constraint $G_J$ enforces physical relations between the quantum numbers $j_e$ of edges $e$ meeting at vertex $v$. The electric part $H_E$ and the magnetic part $H_B$ of the Kogut-Susskind Hamiltonian act on edges and plaquettes, respectively. Physical states may feature multiple excitations $j_e\in\mathbb{N}/2$ on the same edge and can be created by exciting various (potentially overlapping) plaquettes of the lattice. (b) The usual \gls{HEA} enables short and dense gate layers but requires large optimization parameter count. (c) The physics-informed SSP ansatz preserves gauge-invariant excitations (shown for two-edge-plaquettes relevant for the toy model of \cref{Sec5:ToyModel}), thereby reducing the optimization parameter count at the expense of more complicated controlled multi-qubit gates.}
    \label{fig:AlgorithmAnsaetze}
\end{figure}

A promising pathway is quantum simulation on quantum computers \cite{Wozniak2024,Yang2020,Bannuls2019}%,Wiese2021}
. Exploiting the exponential growth of the Hilbert space with qubit count has already proven successful in investigating Abelian gauge theories such as $\mathbb{Z}_2$ \cite{Cochran2025,Mildenberger2022,Schweizer2019} and U(1) \cite{Martinez2016,Banerjee2012,Marcos2013}. However, less progress has been made for non-Abelian gauge theories in $d\geq 2$, despite their physical relevance. This is mainly because the tools established for simulating Abelian theories often do not generalize to the non-Abelian setting. For example, when dealing with $\mathbb{Z}_2$, the vacuum in the strong coupling limit can be prepared with a mean-field ansatz even at large scales \cite{Dusuel2015}. In the non-Abelian setting, however, this ansatz fails due to the bosonic nature of the quantum numbers on each edge, see \cref{fig:AlgorithmAnsaetze}~(a). First works on quantum simulations of SU(2) LGT have focused on truncating the theory to two energy levels per edge \cite{Lee2023EnergyCorrelators,KlcoSavageStryker2020_SU2OneDimQuantum,Calajo2024_SU2IonQudit}, i.e., making it $\mathbb{Z}_2$-like, or simulating the dynamics \cite{Mezzacapo15,Mendicelli23,Zache2023QuantumSpinNetwork}%,Jiang2025}
, often achieved by decomposing the Hamiltonian evolution into sequences of multi-qubit gates known as Trotter decomposition. The large number of required qubits and gates necessitates extremely low error rates, achievable only with future fault-tolerant quantum computers.

Today's \gls{NISQ} computers instead rely on shallow algorithms, such as the \gls{VQE} \cite{McClean2016,Peruzzo2014,Kandala2017}, in which the storage and preparation of the ground state on a quantum device are assisted by a classical optimization procedure. The built-in noise resilience of the \gls{VQE} algorithm at the quantum level comes at the expense of unique challenges caused by sampling many irrelevant states, or the rapidly increasing number of measurements required to reach sufficient accuracy \cite{Bharti2022, Temme2017}. Nonetheless,  \glspl{VQE} on available quantum hardware have already been used to generate accurate results for quantum chemistry with small- to medium-sized molecules \cite{Liegener2024} and in $\mathbb{Z}_2$  LGT \cite{Cochran2025}. The present manuscript demonstrates its viability for non-Abelian gauge theories as well. 

A major challenge in non-Abelian gauge theories is the invariance under their local symmetry group described by the Gauss constraint, which %induces long-range entanglement across the lattice and 
renders most states unphysical. Common \gls{VQE} ans\"atze, such as the {\it hardware-efficient ansatz} (HEA) (see \cref{fig:AlgorithmAnsaetze}~(b)), which samples almost the full Hilbert space, are therefore ill-suited for scaling. Intuitively, this can be explained as follows: at each lattice vertex, the constraint eliminates a fraction $q>1/2$ of the possible states. As the lattice size increases, i.e., the number of vertices $N$ grows, the fraction of physical states among all possible states decreases as $(1-q)^N$. This implies an exponentially unfavorable scaling of HEA: too many irrelevant candidates render HEA impractical for physically relevant lattice sizes.%\footnote{\gls{VQE} in chemistry often deals with constraints by inserting penalty terms in the Hamiltonian, disfavoring states for which the constraint is violated  \cite{Ryabinkin2019}. While easy to implement, it still requires simulating the full Hilbert space.}

To alleviate such scaling issues for the VQE in LGT, in this work we project out (most) unphysical states by expressing the ansatz on the gauge-invariant Hilbert space in the basis of spin-network functions \cite{Penrose1971,RovelliSmolin1995,Baez1996}. %It retains some notion of locality, advantageous to work on a slightly larger Hilbert space. 
This construction, coined {\it systematic state preparation} (SSP) ansatz, only generates gauge-invariant excitations -- effectively excluding the remaining unphysical states. Its core idea is to only insert quantum gates which act on neighboring lattice sites in such a way that the Gauss constraint is preserved, see \cref{fig:AlgorithmAnsaetze}~(c). %In this way, our proposal alleviates scaling issues. 

The paper is structured as follows: in \cref{Sec2:Gauge-Inv} we recall the basics of SU(2) Yang-Mills LGT. We discuss the VQE algorithm and its associated scaling challenges. To alleviate some of these challenges, we establish the SSP protocol. In \cref{Sec5:ToyModel}, we test our framework on a toy model: a single vertex in three spatial dimensions ($d=3$) with periodic boundary conditions, representing the quantization of the symmetry-restricted phase space with imposed translational invariance of classical Yang-Mills theory. Finally, in \cref{Sec6:Implementation}, we use the SSP to determine the vacuum of the toy model in both ideal and emulated noisy environments.
%Emulation of the \gls{VQE}-assisted vacuum-search in both ideal and noisy environments shows the feasibility of the framework even at the state of near-term \gls{NISQ} devices.

\section{Methods}
\label{Sec2:Gauge-Inv}
\subsection{LGT and spin-network representation}

Given a $d\geq2$-dimensional lattice $\gamma$ with spacing $\epsilon$, the discretized dynamics of a Yang-Mills theory \cite{YangMills54} is defined by the \textit{Kogut-Susskind Hamiltonian} \cite{KogutSusskind75} 
\begin{align}\label{eq:KS-Hamiltonian}
    H= g_E\sum_e H_E(e) - g_B \sum_\Box H_B(\Box),
\end{align}
where the electric part $H_E$ acts on the edges $e\in\gamma$ of the lattice and the magnetic part $H_B$ on the plaquettes $\Box$ of $\gamma$, i.e., closed loops composed of four edges. See \cref{app:TheoryDetails} for details. The system is characterized by its coupling parameters $g_E,g_B\geq 0$. The Hamiltonian is invariant under local SU(2) transformations and, as a consequence, only states obeying this symmetry are physical. In the Hamiltonian formulation, this invariance is implemented through the \textit{Gauss constraint}, which generates these local SU(2) transformations and whose discretized version following Guillemin-Sternberg \cite{GuilleminSternberg2013} reads:
\begin{align}\label{eq:Gauss-Constraint}
    G_J(v)= \sum_{e\cap v \neq \emptyset} P_J(e)\;,
\end{align}
where $v$ denotes lattice vertices, i.e., the initial and terminal points, $e(0)$ and $e(1)$, of all edges $e\in\gamma$. The operators $P_J$ are the gauge-covariant fluxes, i.e., the conjugate variables of the holonomies $h$ over the gauge field 
\cite{Balaban1983,AshtekarLewandowski1995,Thiemann2007}. 
These operators act on the Hilbert space of SU(2)-valued functions over the lattice, commonly referred to as the \textit{kinematical Hilbert space}
\begin{align}
    \H= \otimes_{e\in\gamma} \H_e, \quad \H_e = L_2({\rm SU(2)},d\mu)\;,
\end{align}
with $d\mu$ being the Haar measure over SU(2). \Cref{app:TheoryDetails} provides an explicit representation of those operators. In the following, we restrict the kinematical Hilbert space to its physical subspace -- the kernel of the Gauss constraint~\cref{eq:Gauss-Constraint}:
\begin{align}
    \H_G = \{ \psi\in \H \,|\, G_J(v)|\psi\rangle=0 \; \forall v\in\gamma\}\;.
\end{align}
This defines the gauge-invariant Hilbert space. %Restricting to this subspace is necessary to guarantee that the states admit physically meaningful interpretations. 

Several bases of $\H_G$ have been proposed to make computing the ground state feasible, see e.g.~\cite{Andrea24,Cataldi24,Mezzacapo15}. Choosing the correct one is non-trivial, and the consensus has emerged that the most suitable basis often depends on the problem at hand. The following outlines why the spin-network basis is advantageous when having the goal to eventually perform simulations of LGTs on current \gls{NISQ}-type quantum computers.

The spin-network basis for SU(2) relies on the fact that the irreducible representations (irreps) of a compact Lie group form a complete basis for square-integrable functions on the group (according to the Peter\&Weyl theorem \cite{PeterWeyl27}). For SU(2), these irreps are given by the Wigner-$D$-functions, $D^{j_e}_{m_en_e}$, which serve as natural building blocks for states on individual lattice edges \cite{Wigner59}, and are labeled by the quantum numbers
% In the case of SU(2) said irreps correspond to the Wigner-D-functions $D^{(j)}_{mn}$, i.e., \cite{Wigner59}
%\begin{align}\label{eq:PeterWeyl}
%    \mathcal{H}_e\cong \overline{V},\quad V = span (\sum_{jmn} c_{jmn} D^{(j)}_{mn}(h) |c_{jmn}\in\mathbb{C})\;,
%\end{align}
$j_e\in\mathbb{N}/2$ (called spins) and $m_e,n_e\in\{-j,...,j-1,j\}$ for each edge $e$ of the lattice. % Hence, for $\H = \otimes_{e\in\gamma} \H_e $ the basis elements are products of the form $\Pi_{e\in\gamma} D^{j_e}_{m_en_e}$. 
The collection of labels over all edges is denoted by $\vec{j}$, i.e., $j_e\in\vec{j}$ for each edge $e$, and similarly for $\vec{m},\vec{n}$. To enforce gauge invariance, at every vertex the $m,n$-labels of adjacent edges are contracted with SU(2)-invariant tensors called {\it intertwiners}
%Then, $\mathcal{H}$ is projected onto $\mathcal{H}_G$ by contracting, for each tensor product $\psi\in\H$, all magnetic quantum numbers meeting at a vertex $v$ with an SU(2)-invariant intertwiner 
$\iota^v\in \vec{\iota}$, where we again denote the collection of intertwiners for all vertices as $\vec{\iota}$. This gives rise to the spin-network functions \cite{Penrose1971,RovelliSmolin1995,Baez1996}:
\begin{align}
    |T_{\vec{j},\vec{\iota}} \rangle = & \sum_{e_k\in\gamma}\sum_{\vec{m},\vec{n}}(\otimes_{e\in\gamma} |D^{j_e}_{m_e n_e}\rangle ) \\
    \nonumber&( \prod_{v\in\gamma}\iota^v_{m_{e_1}...n_{e_6}}\delta_{e_1(0),v}...\delta_{e_6(1),v})\;\in\cal{H}_G.
\end{align}
Each spin-network function $T_{\vec{j},\vec{\iota}}$ is fully determined by $\vec{j}$ of irreps over all edges $e\in\gamma$ and the choice of intertwiners $\vec{\iota}$ on all vertices $v\in\gamma$. 

For the cubic lattices considered in this work, vertices and their intertwiners are 6-valent. Without loss of generality, 6-valent intertwiners can be decomposed into multiple
%expressed by introducing auxiliary SU(2) irreps and then contracting suitable triples of irreps by coupling them to 
3-valent intertwiners \cite{BrinkSatchler1993}, also known as Wigner-3j-symbols. Those symbols store information about a triple of spins $j$ meeting at the vertex and are denoted by \cite{Wigner59}
\begin{align}
    \iota_{m_1,m_2,m_3} := \left(\begin{array}{ccc}
      j_1 & j_2 & j_3 \\
      m_1 & m_2 & m_3
 \end{array}\right)\;. 
\end{align}
An explicit construction of the decomposition is given in \cref{app:TheoryDetails}. Not all triples of spins $j$ yield non-zero gauge-invariant functions. In other words, the spin-network basis allows for the existence of residual states. These are non-physical states whose associated 3j-symbols automatically vanish unless the following conditions are satisfied \cite{BrinkSatchler1993}:
\begin{align} 
    \label{eq:3j_1_2_condition} &j_1+j_2+j_3\in\mathbb{N}\;,\\
    \label{eq:3j_triangle_condition} |&j_1 - j_2| \leq j_3 \leq j_1+j_2 \;.
\end{align}
Each condition eliminates $1/2$ of the possible combinations of SU(2) irreps at each 3-valent intertwiner. As a consequence, the fraction of physical states for $N$ 3-valent intertwiners decreases as $1/4^N$. Therefore, even in the spin-network basis, randomly sampling the full Hilbert space to find a physical state scales unfavorably.

In summary, while this description of the gauge-invariant Hilbert space reduces a significant amount of complexity compared to simulating the full lattice Hilbert space, it still grows exponentially with lattice size but with a smaller prefactor. Finally, if one restricts to states satisfying \cref{eq:3j_1_2_condition,eq:3j_triangle_condition}, the prefactor reduces further.

%This completes the construction of the gauge-invariant Hilbert space on a cubic lattice via the basis $T_{\vec{j},\vec{\iota}}$. 

\subsection{VQE and its challenges}
\label{Sec4:VQE}

%E.g., determining the ground state of an LGT system benefits from quantum computing methods. Said field is currently in the \gls{NISQ} era, where noise still dominates quantum simulations, making long algorithms impractical \cite{Preskill2018,Liegener2024}. 
 
Projecting onto the spin-network basis allows a lattice to be simulated with fewer qubits than representing the full LGT Hilbert space. Because this reduction directly impacts near-term quantum simulations, we now describe how to facilitate ground state preparation in this setting. Finding the ground state $\Omega$ of the Kogut-Susskind Hamiltonian $H$, i.e., solving
\begin{align}
H |\Omega\rangle = E_0 |\Omega\rangle
\end{align}
on a quantum device, can be achieved using the \gls{VQE} algorithm \cite{McClean2016,Peruzzo2014,Kandala2017}. A \gls{VQE} ansatz $A(\vec{\theta})$ is parametrized by a set of variational parameters $\vec{\theta}$ chosen such that applying $A$ on the digital ground state of a quantum computer, $\ket{0}$, explores a sufficiently large portion of the Hilbert space on which $H$ is defined. If the ansatz is sufficiently expressive, $\Omega$ can be approximated, i.e.,
\begin{align}
\label{eq:VQE_denseness}
\ket{\psi(\vec{\theta})}:=A(\vec{\theta})\ket{0},\quad {\rm s.t.}\; \exists\vec{\theta}^* : \ket{\psi(\vec{\theta}^*)}\approx \ket{\Omega}
\end{align}

\glspl{VQE} have gained significant attention in the last decade due to their shallow circuit depth and variational nature leading to noise resilience. Despite these successes, \glspl{VQE} face significant challenges, especially when scaling the system \cite{Bharti2022, Temme2017}. In particular, the following main challenges can be identified: (i) required number of repeated measurements, (ii) availability of sufficiently many highly-coherent qubits, and (iii) the barren plateau problem for large optimization parameter landscapes. The remainder of this section provides a brief review of these items.

Challenge (i) is related to the exponential growth in the number of required measurements on the system with increasing dimension of $H$: the \Gls{VQE} determines $\vec\theta^*$ using the expectation value of the Hamiltonian as cost function, i.e.,
\begin{align}
    \min_{\vec\theta} \braket{\psi(\vec{\theta}) | H | \psi(\vec{\theta})} \approx E_o\;,
\end{align}
which requires expressing the Hamiltonian in the set $\P_{n_q}$ of all Pauli strings over $n_q$ qubits, such that each string $P\in \P_{n_q}$ corresponds to a single measurement on the quantum device:
\begin{align}
    H = \sum_{\alpha} c_\alpha P_{\alpha}\; .
\end{align}
However, the number of possible qubit strings grows rapidly, $|\P_{n_q}|=4^{n_q}$. If a Hamiltonian has support on all of them, its decomposition would demand a prohibitive number of measurements to obtain sufficient statistics and achieve a low-variance energy estimate.
For example, in typical chemistry simulations the number of Pauli strings grows with $\O(n_q^4)$ \cite{Mathis2020}.

The second challenge (ii) is the scaling of the number of qubits with increasing number of vertices $N$, which is driven by two factors: First, meaningful predictions require small discretization errors, i.e., large lattice sizes. Second, the Wigner functions $D^{(j)}_{mn}$ for SU(2) run over $j\in\mathbb{N}/2$, hence require infinitely many degrees of freedom to be fully captured. It is thus necessary to introduce a cut-off $j_{max}$ to restrict the system to finitely many qubits. Then, the naive qubit count scales as $\mathcal{O}(N j_{max}^{3d})$. %However, only a small portion of this subspace corresponds to physical states, making projection onto the gauge-invariant sector highly advantageous.

Finally, (iii) arises from the trade-off between adding more variational parameters $\vec{\theta}$ (which increase the chance to maximize overlap of the prepared state with the exact ground state as expressed in \cref{eq:VQE_denseness} by sampling more of the Hilbert space) on the one hand and on the other maintaining efficient convergence towards $\theta^*$ (which becomes exponentially harder with increasing number of parameters, also known as the \textit{barren plateau problem} \cite{McClean2016,McClean2018,Wang2021,Cerezo2021}). The scaling of optimization parameters is crucial in settings such as gauge theories, where irrelevant directions exist in the parameter landscape of $\vec{\theta}$ due to the unphysical states and gauge degeneracies inherent in $\H$. This challenge affects all circuits, particularly those featuring large parameter counts (e.g.\@ \glspl{HEA}), which are known to struggle with this trade-off. This represents one of the main challenges \glspl{VQE} need to address.

\subsection{Scaling prospects in LGT with SSP}

To help alleviate the aforementioned challenges, we have designed a new ansatz, coined the {\it systematic state preparation} (SSP), for \glspl{VQE} in non-Abelian LGT and validate its feasibility and accuracy. We summarize its guiding principles here and refer to \cref{app:Circuits} for all details:

SSP utilizes the bosonic nature of the irreps on each link $j\in\{0,1/2,...,j_{max}\}$ combined with the fact that the Hamiltonian only excites closed loops on the lattice. Instead of exciting individual qubits, SSP only generates gauge-invariant excitations by simultaneously manipulating all qubits storing information along said loops. We develop such an ansatz $A$ by utilizing ancilla qubits and more complicated multi-qubit gates, e.g.\@ $A^\pm$ for raising/lowering the quantum numbers $j$. Schematically, we present this in \cref{fig:AlgorithmAnsaetze}~(c), where SSP acts only once per layer on a qubit register, each corresponding to a single irrep $j$. This effectively excludes all the remaining unphysical states and we will argue in the following how it alleviates scaling issues, especially in the context of LGTs.

First, the scaling of Pauli strings is mitigated by both locality and translational invariance of LGTs irrespective of the ansatz used. It was estimated in \cite{Mathis2020} that the number of strings grows roughly with $\O(N\,j_{max}^4)$ with $j_{max}$ being an ad hoc cut-off of the $L_2$-space over each edge to allow encodability using $\log_2(j_{max})$ qubits and $N$ being the number of vertices. %See \cref{app:TheoryDetails} for a brief analysis of the relevancy of scaling $j_{max}$ in a toy model.
For a fixed $j_{max}$, we have $n_q\sim N \log_2(j_{max})$ and the Pauli string scaling is therefore much milder than in, e.g., chemistry. Furthermore, since $[H_E(e),H_E(e')]=0$ for all $e,e'$, a single measurement of all qubits suffices to extract the electric part independent of the lattice size and needs only to be repeated with some measurement count $n_s=n_s(\delta)$ to reach precision $\delta$. Second, $H_B(\Box)$ is local, and if two plaquettes $\Box$ and $\Box'$ do not share an edge, they can be decomposed into Pauli strings and measured simultaneously. Hence, the total magnetic part $\sum_\Box H_B(\Box)$ can be decomposed into $4d$ sets of plaquettes which are pairwise non-intersecting, so the number of required measurements on the quantum device only grows with $\O(j_{max}^4)$. Decoupling the required measurements from lattice size in this way provides a fundamental scaling advantage. However, it does not reduce the measurement count $n_s$ per Pauli string required to reach precision $\delta$, which in uniform allocations still scales with $N$, i.e., $n_s\geq N/\delta^2$ \cite{Wecker2015}. While improved approximations for the required measurement count already exist (e.g.~\cite{zhu2024}), an alternative approach exploits the symmetries of the vacuum in LGT. In particular, the Hamiltonian $H$ (and hence $\Omega$) is invariant under real-space translations $T: v\mapsto v+\epsilon$ that leave the lattice invariant. In other words, it suffices to perform measurements of $H_E(e_o)$ and $H_B(\Box_o)$ at some predefined point on the lattice, given by $e_o$ or $\Box_o$, respectively. This drastically reduces the measurement count, such that it scales with $\O(1/\delta)$ only. However, it comes at the expense of requiring that the prepared states $\psi(\vec{\theta})$ are invariant under $T$. We can in principle achieve this additional condition by ending the SSP ansatz with a gate sequence that performs a \textit{group-average} over all translations \cite{Marolf2000}. The number of operations for symmetrizing a state grows with $\O(N^d)$, thus, this procedure is viable only in the late \gls{NISQ}/fault-tolerant era, where the measurement count becomes prohibitive but gate fidelities are sufficient to extend $A$ in such a way.

Second, the scaling of required qubits with lattice size is alleviated by the fact that SSP projects the Hilbert space to the gauge-invariant space of spin-network functions. Restricting to this set, which is only slightly larger than $\H_G$, allows us to work with a tamer scaling in qubit count. For comparison, when using the basis defined by all Wigner-$D$-functions, $D^{j_e}_{m_en_e}$, one would require a naive scaling in terms of qubit numbers of $n_q\sim n^{\H}_q:=N(3d\log_2(d_{j_{max}})+2d)$ 
%$n_Q\sim\O(N(3d\log_2(d_{j_{max}})+2d))$
with $d_j:=2j+1$. %However, projecting onto the gauge-invariant space reduces this drastically \cite{davoudi2025}
SSP uses explicitly spin-network functions, which carry only one irrep per link but introduce additional intertwiners on each vertex, supported on $\pi\in \{0,...,2j_{max}\}$. This leads to an effective qubit count scaling as  $n_q \sim n^{\H}_q-3N\log_2(d_{j_{max}})+3$ for $d=2,3$, 
%$\O(N((3d-3)\log_2(d_{j_{max}})+(2d-3)))$
which saves qubit resources especially on large lattices, $N\gg 1$, or large cut-offs, $j_{max}\gg 1$.
Using this basis, the quantum computer may also store states whose irrep combinations correspond to trivially vanishing spin-network functions, because they violate \cref{eq:3j_1_2_condition,eq:3j_triangle_condition}. Hence, our construction overestimates $\H_G$ by a small number of states, however, it eliminates most unphysical states\footnote{Recent tensor network approaches advocate the use of the \textit{dressed-site basis} \cite{Magnifico24,Cataldi23}. Here, the remaining gauge conditions are on the edges instead of being at the vertices. The required qubit count scales as $n_q \sim (4d-3)/(3d) n^{\H}_q-(4d-3)$
%$\O(N((4d-3)\log_2(d_{j_{max}})-(2d-3)))$
for the same accuracy, which motivates us to focus on the spin-networks in this manuscript, where the regime $j_{max}>1/2$ is targeted.}. Moreover, the remaining conditions enable easy-to-implement symmetry verification and postselection protocols that aid error mitigation, see \cref{app:SymmVerif} and \cref{Sec6:Implementation}.

Finally, a main motivation of introducing the SSP ansatz is to counteract the barren plateau problem: Reducing to $\H_G$ in itself alleviates this issue, and a further potential is leveraged by replacing the usual \gls{HEA} with a physically inspired ansatz: SSP requires fewer optimization parameters at the expense of an increased number of 2-qubit gates per layer by a factor depending on $d_{j_{max}}$. Crucially, the number of optimization parameters scales as $\O(N)$ per layer instead of as $\O(N j_{max})$, as is the case for the \gls{HEA}. Given that barren plateau problems arise when gradients vanish exponentially fast with the number of optimization parameters \cite{McClean2018,Holmes2022}, this reduction provides a significant speed-up, as will be demonstrated in the next section.

\section{SSP performance on the 6-vertex toy model}
In this section, we test the SSP ansatz by applying it to a toy model of 3D Yang-Mills theory with SU(2) gauge group, which can be analyzed with emulators of mid-scale quantum devices. 

\subsection{Model: 6-valent vertex}
\label{Sec5:ToyModel}

The model consists of a single vertex $v$ with periodic boundary conditions. Conceptually, we interpret this as a homogeneous system in which translation-invariance has been imposed prior to quantization. Note that, on the classical level, imposing translational invariance does not affect the symmetry-restricted phase space and its dynamics \cite{KaminskiLiegener2020}. Yet, the extension of this phenomenon to the quantum level is still under investigation.

%\footnote{Note that translation invariance acts fundamentally differently on classical and quantum space. While the homogeneous configuration can be described by a single degree of freedom classically, the same is not true for the quantum Hilbert space: entanglement between different sites may persist even after imposing translation invariance.}
In our case, given that Hamiltonian and ground state are both translation invariant, we can still capture some relevant properties of a full 3D LGT in this reduced setting, such as correlations between different spatial directions: As many contemporary works focus on 2D simulations, their findings could be impacted by the presence of correlations between all three dimensions as this model allows to investigate. Further, since it can be fully simulated classically, it allows us to benchmark the accuracy of the VQE results.

\begin{figure}
    \centering
    \includegraphics[width=1\linewidth]{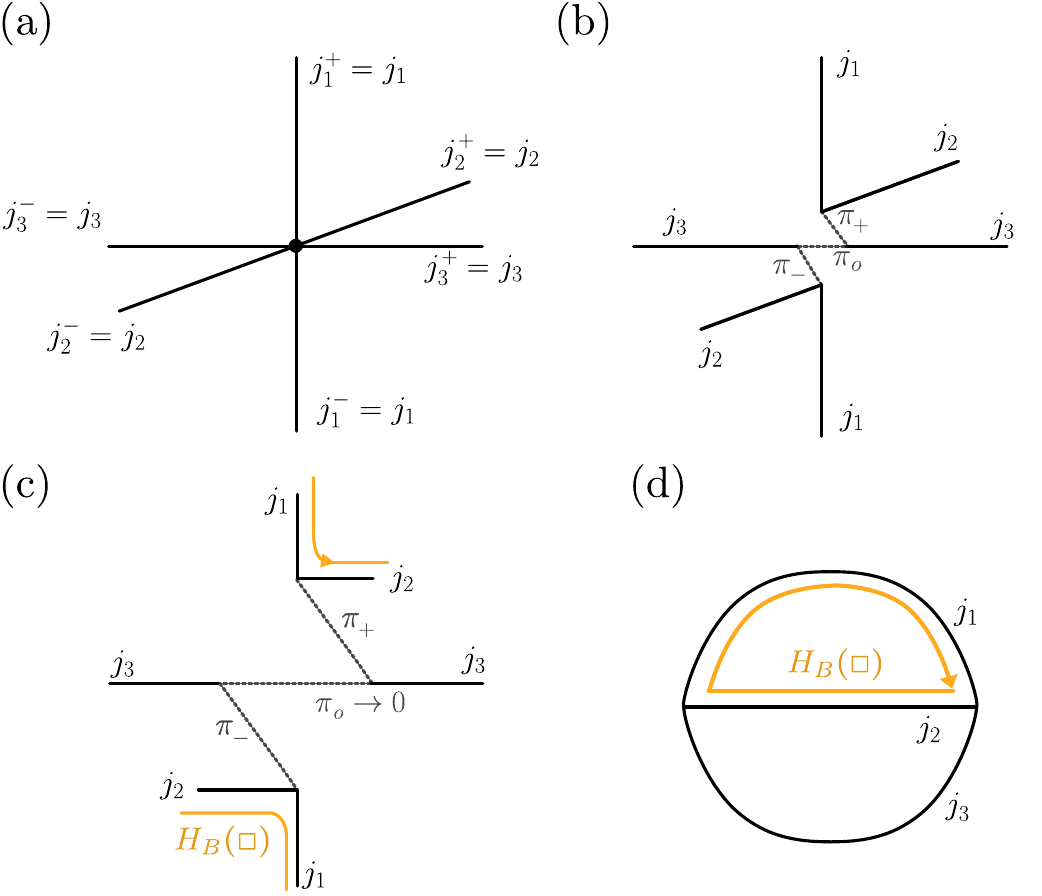}
    \caption{Graphical representation of the toy model. (a) A single vertex in 3D with periodic boundary conditions, i.e., in- and outgoing irreps in each direction carry the same label. Projecting to the gauge-invariant Hilbert space is accomplished by choosing a basis of intertwiners $(\pi_+,\pi_o,\pi_-)$ in (b). (c) We choose a discretization of the Hamiltonian approximating continuum Yang-Mills by attaching plaquettes such that $\pi_o$ is not changed by action of $H_B$, hence separating the graph into superselection sectors. (d) Focusing on the sector $\pi_o=0$ reduces the single-vertex graph to the $\Theta$-graph.}
    \label{fig:1VertexGraph}
\end{figure}

By imposing periodic boundary conditions on the incoming and outgoing edges of the vertex, we simplify the Hilbert space to three copies over SU(2), one for each direction, labeled by irreps $j^{\pm}_k:=j_k$, $k\in\{1,2,3\}$. Going to the gauge-invariant Hilbert space contracts the corresponding Wigner-D functions with an intertwiner $\iota^{v,\pi_+,\pi_o,\pi_-}$ as specified in \cref{app:TheoryDetails} in equation \cref{eq:intertwiner3D}, characterized by the internal quantum numbers $\pi_-,\pi_o,\pi_+$. We discretize the Hamiltonian on this graph as follows: $H_E(e)$ as in \cref{eq:KS-Hamiltonian} acts on the unique edge for each direction. We note that $H_B(\Box)$ stems from a discretization of the curvature of the connection at $v$. To approximate it fully, it suffices at each vertex to consider only the plaquettes $\Box_{kl}$ oriented along pairs of positive outgoing edges $j^+_k,j^+_l$. Due to the periodic boundaries, the minimal plaquettes encompass only two edges intersecting $v$ and still approximate the curvature. This orientation choice simplifies the action of $H$ as in the chosen basis for $\iota^{v,\pi_+,\pi_o,\pi_-}$, the label $\pi_o$ remains invariant under the action of such a discretized $H$. Thus, the Hilbert space decomposes into superselection sectors with respect to the Hamiltonian, with each eigenvector supported in only one sector labeled by $\pi_o$. %Classical evaluation confirms that t
The ground state lies in the sector $\pi_o=0$, which -- due to the symmetries of the Wigner 3j-symbols -- enforces $\pi_\pm=j_3^\pm=j_3$. In summary, this superselection sector of $\H$ is entirely characterized by $j_1,j_2,j_3$ and hence becomes isomorphic to the Hilbert space over a $\Theta$-graph, see \cref{fig:1VertexGraph}. See \cref{app:TheoryDetails} for further details.

We stress that the Hilbert space over this finite lattice is still infinite dimensional ($j_k\in {\mathbb N}/2\;\forall k$) and for it to be stored on a quantum device with finite qubit count, a cut-off $j_{max}$ is required. This implies an unavoidable error in the full model, which requires careful analysis to be faithfully estimated. Yet, while recently proposed methods estimate this error for Abelian gauge groups \cite{Ciavarella2025}, none have been extended to the non-Abelian setting so far. In this specific model, we can use the error introduced by cut-off $j_{max}$ to gauge the accuracy of the \gls{VQE} results, as deviations of the energy estimates below said error are no longer physically relevant. %\footnote{It is worthwhile to see that the relevance of the model drastically increases by raising the cut-off from $j_{max}=1/2$ (the analog to the customary work in $\mathbb{Z}_2$ LGT) to $j_{max}=3/2$ as is done in this manuscript. In turn, further accuracy is achieved by raising the cut-off further, see \cref{app:TheoryDetails}.}
 In other words, we benchmark the effectiveness of the proposed SSP ansatz by how closely its energy estimates fall within the uncertainty margin introduced by the finite cut-off. This is particularly relevant in the strongly correlated regime, where the ground state no longer factorizes over the edges but is driven by non-local entanglement.

\subsection{Simulated Implementation}
\label{Sec6:Implementation}

To demonstrate that the SSP ansatz accurately estimates energy and identifies distinct phases of spatial correlation, we simulate its implementation.

The configuration space of coupling constants is described by the open set $g_E,g_B\geq 0$, which we compactify to a finite closed interval: Since any quantum system is invariant under a total reparametrization of the energy, only the total ratio between both couplings is physically relevant.\footnote{Note that the Yang-Mills action \cite{YangMills54} is defined by a single coupling constant. The second one emerges upon discretization when the Hamiltonian depends on the lattice spacing $\epsilon$ \cite{KogutSusskind75}. Recovery of the continuum requires the limit $\epsilon\to0$ and a running of the coupling constants in form of a renormalization-group flow \cite{lang2017hamiltonianI}. Here, we refrain from this extension and do not associate the coupling parameters with their continuum values.} Hence, we express both in terms of a single parameter $g_E=:(1-\lambda^2)$ and $g_B=:2\lambda^2$ with $\lambda\in[0,1]$, as both strong and weak coupling regimes are recovered in the limits $\lambda\to 0$ and $\lambda\to 1$, respectively. %Note that especially the latter limit in which $H_B$ dominates is of interest for non-Abelian gauge theory, where the ground state in the $\lambda\to 1$ limit is not an analytically well-known state -- in contrast to $\mathbb{Z}_2$ LGT where most contemporary work only looks at finite segments of the configuration space, e.g.~\cite{Cochran2025}.
This compactification is especially valuable for the following analysis, where the \gls{VQE} minimization of the ground state for the SU(2) Yang-Mills  toy model now samples the whole configuration space.

We simulate the described SSP ansatz from \cref{Sec4:VQE} for the \gls{VQE} first in an ideal setting without the coupling to the environment. The SSP variants of different depths are explicitly constructed in \cref{app:Circuits}, especially for the cut-off $j_{max}=3/2$. In this section, we restrict to this cut-off, in which case the quantum number for each edge can be expressed with two qubits, i.e., $\ket{q_{k,1}, q_{k,2}} \equiv j_k\in\{0,1/2,1,3/2\}$. Note that SSP requires up to two ancilla qubits (or more for even deeper variants than the ones used here), meaning that the total quantum resources for our toy model amount to 8 qubits (= 6 data + 1 or 2 ancillas). A more detailed overview of the key properties for each SSP and \gls{HEA} is presented in table \cref{table:KPI} in \cref{app:SymmVerif}. In our simulations, we employed the Powell optimization method due to its robustness to noise \cite{powell1964_powell,virtanen2020scipy_powell}, and adjusted the count of randomized starting candidates to increase linearly with the optimization-parameter count.

We start by comparing the performance of the newly introduced SSP with the conventional HEA. Note that both refer to families of ans\"atzen, e.g.\@ we label SSP by the
different counts $n=2,3,4$ of gauge-invariant plaquette excitations to be added while preparing the ground state, calling the respective variant SSP$n$. Conversely, HEA$m$ denotes a variant with $m$ many optimization parameters. With increasing optimization parameter count, equivalent to deeper circuits, the variants will sample larger portions of the Hilbert space. Consequently, it will always be possible to find within both families a variant that prepares a state $\psi(\theta^*_\lambda)$ approximating the ground state $\Omega$ better than a targeted infidelity

\begin{figure}
    \centering
    \includegraphics[width=1\linewidth]{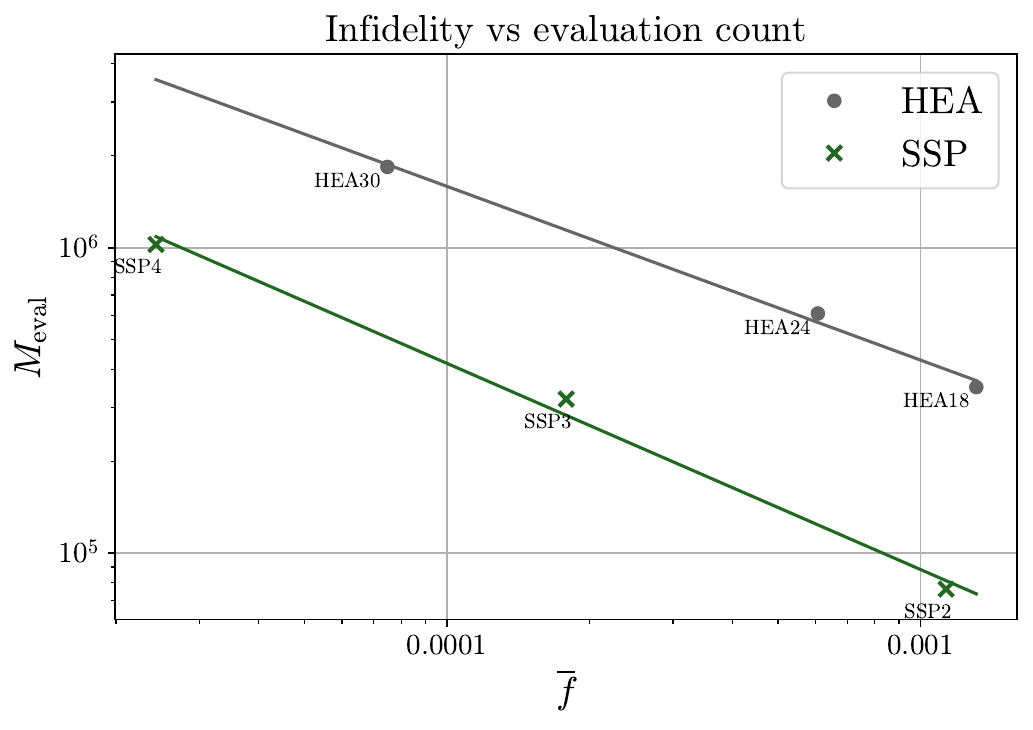}
    \caption{Comparison of HEA (grey dots) and SSP (green crosses). Achieving a given infidelity $\bar f$ (averaged over all $\lambda$) requires deeper circuits with more parameters and implies more function evaluations $M_{\rm eval}$, i.e., calls to the quantum computer. For both ans\"atze, a fit (linear on the log-log scale) has been added as $M_{\rm eval}(\bar f)=a\; \exp(b \bar f)$ with fit parameters $(a,b)=(-0.5, 9)$ for HEA and $(a,b)=(-0.6,6.7)$ for SSP respectively.   
    }
    \label{fig:scaling_funcevals}
\end{figure}

\begin{align}
    \bar f = \int_0^1 d\lambda \; f_\lambda,\quad f_\lambda = 1 - |\braket{\Psi(\theta^*_\lambda) | \Omega_\lambda}|^2,
\end{align}
averaged over all couplings $\lambda$. While low $\bar f$ establishes suitable accuracy in applications, we must further ask how costly it is to reach it given an ansatz. Assuming fault-tolerant quantum computing, the limiting factor is the count of function evaluations, $M_{\rm eval}$ (i.e., the count of how often the expectation value has to be computed during the classical optimization procedure). Note that we define $M_{\rm eval}$ as the average number of function evaluations needed for the optimizer to converge to a local minimum given an arbitrary starting location. In fact, a mild scaling of $M_{\rm eval}$ is necessary for any \gls{VQE} application to be feasible. In \cref{fig:scaling_funcevals}, we compare the two families SSP and HEA to see how many evaluations $M_{\rm eval}$ are necessary to be better than a certain infidelity $\bar f$. Both (HEA18, HEA24, HEA30) and (SSP2, SSP3, SSP4) achieve lower infidelities $\bar f$ with increasing number of optimization parameters, which requires more function evaluations (see \cref{table:KPI} in \cref{app:SymmVerif} for details). As expected from the barren plateau problem, the dependence on evaluation count is exponentially related to the number of optimization parameters as can be seen in the fitted exponential decay $M_{\rm eval}(\bar f)$ in \cref{fig:scaling_funcevals}. To summarize: in the investigated regime, SSP outperforms HEA in terms of function evaluations. %\footnote{The simulation assumed ideal 2-qubit gates. Deeper algorithms will help to further decrease infidelity; however, they require fault-tolerant devices, which are not yet accessible. Due to the error bars of numerical optimization, an estimation beyond the simulated regime is not reliable.}

\begin{figure}
    \centering
    \includegraphics[width=1\linewidth]{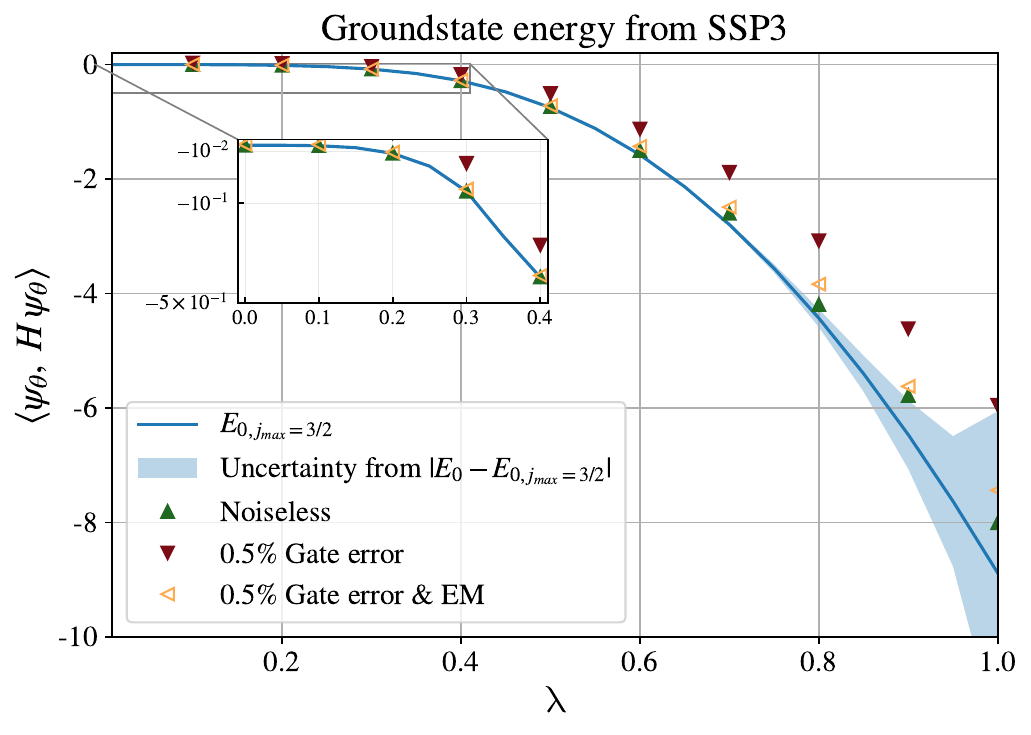}
    \caption{Best achieved minima for the energy expectation values for SSP3, a particular variant of the SSP circuit which  creates $n=3$ gauge-invariant plaquette excitations. 
    The \gls{VQE} has been emulated in an ideal environment (green) and in the presence of noise from a 0.5\% 2-qubit gate error (red). Upon adding error mitigation (EM) protocols (orange), the accuracy comes close to the exact energies (blue) and falls within the uncertainty given by the finite cut-off $j_{max}=3/2$ (shaded region).}
    \label{fig:energy_exp_val_ssp3}
\end{figure}

As realistic quantum devices in the \gls{NISQ} era are never noise-free, we take a closer look at the SSP3 variant and investigate its performance when coupled to the environment (see \cref{app:DigitalTwin} and \cref{app:HEA_VQE} for more details on the other circuits). In \cref{fig:energy_exp_val_ssp3}, we present the best energy estimates for the ground state obtained from the \gls{VQE} runs: In an idealized noise-free setting, for low $\lambda$ values the short SSP3 circuit suffices, as the green triangles are close to the exact numerically determined solution $E_{0,j_{max=3/2}}$ shown in blue and computed in \cref{app:TheoryDetails}.
%shorter circuits suffice, as one can see that SSP2 -- bullet points in (a) -- perform similar to  SSP3 -- triangles in (b) -- and SSP4 -- squares in (c).
However, towards $\lambda\to 1$, the energies of the prepared ground states differ from the exact values, confirming that the true ground state requires more gauge-invariant excitations to approximate the analytically known solution at $\lambda=1$, i.e., a Dirac-$\delta$ distribution centered at the unit elements of SU(2).
%\footnote{This originates from the analytic ground state at $\lambda=1$ corresponding to a Dirac-$\delta$ distribution centered at the unit elements of SU(2) which is delocalized over the electric field basis. Nonetheless, even at finite $j_{max}$ the expectation values converge quickly towards the correct vacuum energy.}
%In this ideal setting, SSP4 achieves sufficient accuracy to qualitatively describe the 6-valent vertex model with cut-off $j_{max}=3/2$.
%For comparison, in blue are the exact values of the ground state computed in \cref{app:TheoryDetails}, including an analysis of the impact of cut-off choice. 

Further, we simulate the LGT model by emulating the noise of contemporary quantum computers as explained in detail in \cref{app:DigitalTwin}, built with a square lattice topology and such that each 2-qubit gate is assigned a $0.5\%$ error. The presence of noise leads to an overall loss of accuracy, especially in the large $\lambda$ limit. However, these noisy results, presented in red in \cref{fig:energy_exp_val_ssp3}, should afterwards be processed with error mitigation tools. We outline in \cref{app:SymmVerif} the post-selection protocols applied, which can be used during the optimization procedure: projecting the prepared state back to the gauge-invariant Hilbert space and imposing rotational symmetry. At this stage, we are interested only in the best values from the minimization runs to determine the best-fit optimization parameters.\footnote{Conceptually, re-running a state preparation algorithm with the same setting gives spread-out results due to the presence of noise, which affects the found optimum in a single run -- but these effects are not taken into account in \cref{fig:energy_exp_val_ssp3}, which presents a single optimization run.}
While the \gls{VQE} method targets the Hamiltonian at cut-off $j_{max}=3/2$, shown in solid blue in \cref{fig:energy_exp_val_ssp3}, the shaded area in light blue stems from the uncertainty due to the necessary cut-off error. Here, we computed the cut-off error exactly, yet future applications on larger systems need to replace this exact value by suitable approximations building upon \cite{Ciavarella2025}.

\subsection{Measurement of 2-point correlators}

\begin{figure}[ht]
    \centering
    \includegraphics[width=1\linewidth]{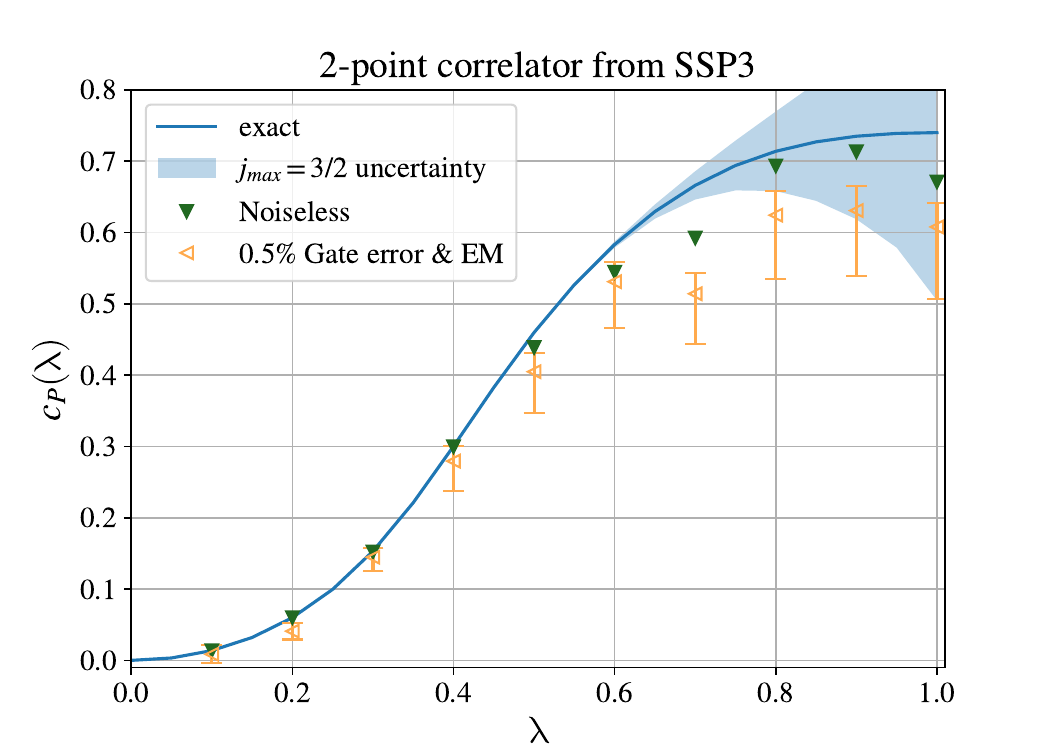}
    \caption{2-point correlator evaluated on $\psi(\vec\theta^*)$ prepared with the best optimization parameters found in \cref{fig:energy_exp_val}. Green triangles show ideal values, and orange points show error-mitigated and post-selected results. Data points indicate median and error bars indicate standard deviation centered around the mean. In blue is the exact correlator shown with a shaded uncertainty region due to the finite cut-off $j_{max}=3/2$.}
    \label{fig:correlator}
\end{figure}

Given the optimal parameter set $\vec{\theta}^*$, the SSP method creates the target state $\psi(\vec{\theta}^*)$ close to the exact ground state $\Omega$ of the toy model. Using this state, we prepare the vacuum to study physical properties of the model. Primarily, we aim to analyze the different phases of the LGT model. To distinguish these from each other, we need to introduce a suitable gauge-invariant observable. We outline in \cref{app:TheoryDetails} how the toy model of the 6-valent vertex naturally gives rise to a gauge-invariant, discretized version of the 2-point correlator, symbolically written to as
$\langle \Omega | (A^I_a A^J_b)^G|\Omega\rangle$. Once the ground state $\Omega$ of the theory has been prepared for a given coupling parameter $\lambda$, we use it to measure the expectation value of said 2-point correlator. Analyzing these correlations allows one to identify the coupling-dependent phases of the model. Based on symmetry arguments, the non-vanishing terms of the 2-point correlator of the toy model must be of the form  (see \cref{app:TheoryDetails} for details)
\begin{align}
    \langle\Omega | (A^I_a A^J_b)^G | \Omega\rangle = \delta_{IJ} c_P(\lambda)
\end{align}
where $a\neq b$ and $c_P(\lambda)$ is obtained by evaluating the expectation value of a Pauli decomposition of that operator. The 2-point correlator is shown in \cref{fig:correlator}. Here, the best optimized $\vec{\theta}^*$ from the simulations shown in \cref{fig:energy_exp_val_ssp3} have been used to prepare the ground state with a sample size of $1,000$ and the median shown as orange data points cleaned in post-processing analysis as outlined before. The error bars denote the standard deviation centered around the mean of the histogram population. After error mitigation, the different phases for the coupling parameter can be clearly distinguished\footnote{Note that in the presence of a cut-off $j_{max}$, the system is finite-dimensional, thus not exhibiting true ``phase transitions''. We refer to \textit{facsimile} phases in the following, whose properties are expected to hold when taking the corresponding limit $j_{max}\to \infty$.}: the uncorrelated phase defined by $c_p\approx0$ dominates the limit $\lambda\to 0$, and correlations between the different directions of the lattice emerge for large $\lambda\to1$ values, indicated by $c_p\approx {\cal O}(1)$. The system slowly undergoes a gradual crossover (phase transition) for intermediate $\lambda$. In fact, independent of the depth of SSP used for ground state preparation, the crossover is observed in the same region. At the simulated $0.5\%$ gate-error level, the exact correlator values cannot be reached for all $\lambda\in[0,1]$.

\section{Discussion}
\label{Sec7:Conclusion}

We introduced the SSP protocol to investigate ground state preparation via \gls{VQE} on quantum hardware for non-Abelian LGTs.
Although the scalability of \gls{VQE} in general remains an open question, we alleviate some of its challenges, e.g.\@ the barren plateau problem, by using the SSP ansatz. Its potential for scaling to larger lattice sizes rests on projecting onto the spin-network basis, thereby reducing the required qubit resources for representing the system. Additionally, by preserving gauge invariance during creation of the ground state, it reduces the optimization parameter count. The reduction of required optimization parameters makes the SSP ansatz particularly attractive for the late \gls{NISQ} era, where qubit numbers and gate fidelities have matured to a level at which \gls{VQE} is limited by the barren plateau problem. 

We demonstrated the feasibility of SSP in a classically symmetry-restricted $d=3$ pure SU(2) Yang-Mills toy model consisting of a single 6-valent vertex. In an ideal setting, SSP achieves arbitrary accuracy by successively increasing circuit depth. Even in the presence of the noise expected from contemporary quantum hardware, the usual error-mitigation schemes can help to restore accuracy. A comparison between SSP and HEA unveiled that SSP needs consistently fewer function evaluations to converge, thereby alleviating the barren plateau problem. As this is tied to the smaller count of optimization parameters for SSP compared to HEA, we expect this to hold also when scaling to larger lattice sizes, where the fraction of physical states tilts further in favor of SSP. Finally, the prepared ground state was used to extract physical properties, such as the 2-point correlator, which qualitatively detects phase transitions directly from measurements on the quantum device.

%With improved hardware, even deeper SSP circuits than those presented here will become feasible. Given the rapid development in the field, we anticipate that sub-percentage error gates will soon become available, and we reserve experimental verification of our results for future research.

To go further and use the SSP ansatz in regimes beyond classical simulatability, more work is required: to avoid an unfavorable scaling in the number of optimization candidates, one should add an adiabatic ramp-up protocol for large lattices. Additionally, the iterative nature of inserting more excitations into the system until convergence is reached is beneficial in conjunction with the scalable ADAPT-VQE framework, which dynamically adapts the algorithm during the optimization procedure \cite{Grimsley2019}. Finally, moving from the ancilla-based approach to using the plaquette excitations as basic excitations in a unitary coupled cluster ansatz \cite{OMalley2016UCCSD}  can make the algorithm more dense. In summary, a straighforward path to extending SSP is apparent.

Once such states are prepared, one can place initial matter configurations on it to study dynamical evolution \cite{Cochran2025} or obtain insights into the mass gap, e.g., by scaling the system to larger sizes and investigating its asymptotic behavior. %Comparing these and other observables across different lattice sizes will ultimately become a crucial tool in determining suitable Hamiltonian renormalization group maps whose fixed points are required for reliable continuum predictions \cite{lang2017hamiltonianI}. 
We reserve these topics for future research.

\section*{Acknowledgements}
We thank Jad Halimeh, Giovanni Cataldi, Emanuele Mendicelli, Saeed Rastgoo, Jorden Roberts, Federica Fragomeno, Philipp Aumann, and Arya Suneesh for insightful discussions and helpful comments at various stages of this article. We are grateful for many discussions with the QC4HEP consortium. This work received financial support from the German Federal Ministry of Education and Research via the funding program under contract number 13N16188 (MUNIQC-SC) and under contract number 13N15680 (GeQCoS), as well as by the European Union by the EU Flagship on Quantum Technology HORIZON-CL4-2022-QUANTUM-01-SGA project 101113946 OpenSuperQPlus100 and MOlecular Quantum Simulations (MOQS) (Grant-Nr. 955479). The research is part of the Munich Quantum Valley, which is supported by the Bavarian state government with funds from the Hightech Agenda Bayern Plus initiative.

\bibliography{references.bib}

@misc{JaffeWitten2000-YangMillsMassGap,
  author       = {Arthur Jaffe and Edward Witten},
  title        = {Yang--Mills Existence and Mass Gap},
  howpublished = {Millennium Prize Problems, Clay Mathematics Institute},
  year         = {2000},
  url          = {https://www.claymath.org/millennium/yang-mills-and-mass-gap},
  note         = {Accessed: 2025‑08‑30}
}

@article{Balaban1984I,
  author  = {T. Balaban},
  title   = {Propagators and Renormalization Transformations for Lattice Gauge Theories. I},
  journal = {Communications in Mathematical Physics},
  volume  = {95},
  pages   = {17--40},
  year    = {1984}
}

@article{Balaban1983,
  author    = {Tadeusz Balaban},
  title     = {Renormalization Group Approach to Lattice Gauge Theories. I. Generation of Effective Actions in a Small Field Approximation and a Coupling Constant Renormalization in Four Dimensions},
  journal   = {Communications in Mathematical Physics},
  year      = {1983},
  volume    = {89},
  pages     = {571--597},
  doi       = {10.1007/BF01206135}
}

@article{PhysRevLett.123.042002,
  title = {Quark-Mass Dependence of Elastic $\ensuremath{\pi}K$ Scattering from QCD},
  author = {Wilson, David J. and Brice\~no, Ra\'ul A. and Dudek, Jozef J. and Edwards, Robert G. and Thomas, Christopher E.},
  collaboration = {for the Hadron Spectrum Collaboration},
  journal = {Phys. Rev. Lett.},
  volume = {123},
  issue = {4},
  pages = {042002},
  numpages = {7},
  year = {2019},
  month = {Jul},
  publisher = {American Physical Society},
  doi = {10.1103/PhysRevLett.123.042002},
  url = {https://link.aps.org/doi/10.1103/PhysRevLett.123.042002}
}

@article{FontanaTrombettoni2025,
  author       = {Pierpaolo Fontana and Andrea Trombettoni},
  title        = {Mean Field Approaches to Lattice Gauge Theories: A Review},
  journal      = {Entropy},
  year         = {2025},
  volume       = {27},
  number       = {3},
  pages        = {250},
  doi          = {10.3390/e27030250},
  url          = {https://doi.org/10.3390/e27030250}
}

@article{BrowerChrist2018LatticeQCDExascale,
  author       = {Richard Brower and Norman Christ and Carleton DeTar and Robert Edwards and Paul Mackenzie},
  title        = {Lattice QCD Application Development within the US DOE Exascale Computing Project},
  journal = {EPJ Web of Conferences},
  volume       = {175},
  pages        = {09010},
  year         = {2018},
  doi          = {10.1051/epjconf/201817509010},
  url          = {https://doi.org/10.1051/epjconf/201817509010}
}

@article{PhysRevLett.112.201601,
  title = {Tensor Networks for Lattice Gauge Theories and Atomic Quantum Simulation},
  author = {Rico, E. and Pichler, T. and Dalmonte, M. and Zoller, P. and Montangero, S.},
  journal = {Phys. Rev. Lett.},
  volume = {112},
  issue = {20},
  pages = {201601},
  numpages = {5},
  year = {2014},
  month = {May},
  publisher = {American Physical Society},
  doi = {10.1103/PhysRevLett.112.201601},
  url = {https://link.aps.org/doi/10.1103/PhysRevLett.112.201601}
}

@article{SilviRicoCalarcoMontangero2014,
  author       = {P. Silvi and E. Rico and T. Calarco and S. Montangero},
  title        = {Lattice gauge tensor networks},
  journal      = {New Journal of Physics},
  volume       = {16},
  pages        = {103015},
  year         = {2014},
  doi          = {10.1088/1367-2630/16/10/103015},
  url          = {https://arxiv.org/abs/1404.7439}
}

@article{Verstraete2008,
  author  = {F. Verstraete and V. Murg and J. I. Cirac},
  title   = {Matrix Product States, Projected Entangled Pair States, and variational renormalization group methods for quantum spin systems},
  journal = {Advances in Physics},
  volume  = {57},
  pages   = {143--224},
  year    = {2008},
  doi     = {10.1080/14789940801912366},
  url     = {https://doi.org/10.1080/14789940801912366}
}

@article{Bannuls2019,
  author  = {M. C. Bañuls and Krzysztof Cichy},
  title   = {Review on Novel Methods for Lattice Gauge Theories},
  journal = {Reports on Progress in Physics},
  volume  = {83},
  pages   = {024401},
  year    = {2020},
  doi     = {10.1088/1361-6633/ab6311},
  url     = {https://arxiv.org/abs/1910.00257}
}

@article{Yang2020,
  author  = {Bing Yang et al.},
  title   = {Observation of Gauge Invariance in a 71-Site Bose–Hubbard Quantum Simulator},
  journal = {Nature},
  volume  = {587},
  pages   = {392–396},
  year    = {2020},
  doi     = {10.1038/s41586-020-2910-8},
  url     = {https://www.nature.com/articles/s41586-020-2910-8}
}

@article{Wozniak2024,
  author  = {C. A. Wozniak et al.},
  title   = {Quantum Computing for High-Energy Physics: State of the Art and Challenges},
  journal = {PRX Quantum},
  volume  = {5},
  pages   = {037001},
  year    = {2024},
  doi     = {10.1103/PRXQuantum.5.037001},
  url     = {https://link.aps.org/doi/10.1103/PRXQuantum.5.037001}
}

@article{Mildenberger2022,
  author  = {Julius Mildenberger and Wojciech Mruczkiewicz and Jad C. Halimeh and Zhang Jiang and Philipp Hauke},
  title   = {Probing Confinement in a \(\mathbb{Z}_2\) Lattice Gauge Theory on a Quantum Computer},
  journal = {arXiv preprint},
  year    = {2022},
  eprint  = {2203.08905},
  archivePrefix = {arXiv},
  primaryClass = {quant-ph},
  url     = {https://arxiv.org/abs/2203.08905}
}

@article{Schweizer2019,
  author  = {Christian Schweizer and Fabian Grusdt and Moritz Berngruber and Luca Barbiero and Eugene Demler and Nathan Goldman and Immanuel Bloch and Monika Aidelsburger},
  title   = {Floquet approach to \(\mathbb{Z}_2\) lattice gauge theories with ultracold atoms in optical lattices},
  journal = {Nature Physics},
  volume  = {15},
  pages   = {1168--1173},
  year    = {2019},
  doi     = {10.1038/s41567-019-0649-7},
  url     = {https://www.nature.com/articles/s41567-019-0649-7}
}

@article{Martinez2016,
  author  = {E. A. Martinez and C. A. Muschik and P. Schindler and D. Nigg and A. Erhard and M. Heyl and P. Hauke and M. Dalmonte and T. Monz and P. Zoller and R. Blatt},
  title   = {Real-time dynamics of lattice gauge theories with a few-qubit quantum computer},
  journal = {Nature},
  volume  = {534},
  pages   = {516--519},
  year    = {2016},
  doi     = {10.1038/nature18318},
  url     = {https://arxiv.org/abs/1605.04570}
}

@article{Banerjee2012,
  author  = {Debasish Banerjee and Markus Dalmonte and Matthias Müller and Enrique Rico and Peter Stebler and U.-J. Wiese and Peter Zoller},
  title   = {Atomic Quantum Simulation of Dynamical Gauge Fields Coupled to Fermionic Matter},
  journal = {Physical Review Letters},
  volume  = {109},
  pages   = {175302},
  year    = {2012},
  doi     = {10.1103/PhysRevLett.109.175302},
  url     = {https://journals.aps.org/prl/abstract/10.1103/PhysRevLett.109.175302}
}

@article{Marcos2013,
  author  = {David Marcos and Philipp Rabl and Enrique Rico and Peter Zoller},
  title   = {Superconducting Circuits for Quantum Simulation of Dynamical Gauge Fields},
  journal = {Physical Review Letters},
  volume  = {111},
  pages   = {110504},
  year    = {2013},
  doi     = {10.1103/PhysRevLett.111.110504},
  url     = {https://arxiv.org/abs/1306.1674}
}

@article{Liegener2024,
  author       = {Liegener, Klaus and Morsch, Oliver and Pupillo, Guido},
  title        = {Solving quantum chemistry problems on quantum computers},
  journal      = {Physics Today},
  volume       = {77},
  number       = {9},
  pages        = {34--42},
  month        = sep,
  year         = {2024},
  publisher    = {American Institute of Physics},
  doi          = {10.1063/pt.qoys.tiuw},
  url          = {https://doi.org/10.1063/pt.qoys.tiuw}
}

@article{KogutSusskind75,
  title = {Hamiltonian formulation of Wilson's lattice gauge theories},
  author = {Kogut, John and Susskind, Leonard},
  journal = {Phys. Rev. D},
  volume = {11},
  issue = {2},
  pages = {395--408},
  numpages = {0},
  year = {1975},
  month = {Jan},
  publisher = {American Physical Society},
  doi = {10.1103/PhysRevD.11.395},
  url = {https://link.aps.org/doi/10.1103/PhysRevD.11.395}
}

@article{YangMills54,
  title = {Conservation of Isotopic Spin and Isotopic Gauge Invariance},
  author = {Yang, C. N. and Mills, R. L.},
  journal = {Phys. Rev.},
  volume = {96},
  issue = {1},
  pages = {191--195},
  numpages = {0},
  year = {1954},
  month = {Oct},
  publisher = {American Physical Society},
  doi = {10.1103/PhysRev.96.191},
  url = {https://link.aps.org/doi/10.1103/PhysRev.96.191}
}

@book{GuilleminSternberg2013,
  author    = {V. Guillemin and S. Sternberg},
  title     = {Semi-classical Analysis},
  publisher = {International Press},
  year      = {2013}
}

@article{Thiemann2007,
  author    = {T. Thiemann},
  title     = {Modern Canonical Quantum General Relativity},
  journal = {Cambridge University Press},
  year      = {2007}
}

@article{AshtekarLewandowski1995,
  author    = {Abhay Ashtekar and Jerzy Lewandowski},
  title     = {Projective Techniques and Functional Integration for Gauge Theories},
  journal   = {Journal of Mathematical Physics},
  year      = {1995},
  volume    = {36},
  number    = {5},
  pages     = {2170--2191},
  doi       = {10.1063/1.531037}
}

@article{Cataldi24,
  title = {Simulating $(2+1)\mathrm{D}$ SU(2) Yang-Mills lattice gauge theory at finite density with tensor networks},
  author = {Cataldi, Giovanni and Magnifico, Giuseppe and Silvi, Pietro and Montangero, Simone},
  journal = {Phys. Rev. Res.},
  volume = {6},
  issue = {3},
  pages = {033057},
  numpages = {23},
  year = {2024},
  month = {Jul},
  publisher = {American Physical Society},
  doi = {10.1103/PhysRevResearch.6.033057},
  url = {https://link.aps.org/doi/10.1103/PhysRevResearch.6.033057}
}

@article{Andrea24,
  title = {New basis for Hamiltonian SU(2) simulations},
  author = {D'Andrea, Irian and Bauer, Christian W. and Grabowska, Dorota M. and Freytsis, Marat},
  journal = {Phys. Rev. D},
  volume = {109},
  issue = {7},
  pages = {074501},
  numpages = {32},
  year = {2024},
  month = {Apr},
  publisher = {American Physical Society},
  doi = {10.1103/PhysRevD.109.074501},
  url = {https://link.aps.org/doi/10.1103/PhysRevD.109.074501}
}

@article{Mezzacapo15,
  title = {Non-Abelian SU(2) Lattice Gauge Theories in Superconducting Circuits},
  author = {Mezzacapo, A. and Rico, E. and Sab\'{\i}n, C. and Egusquiza, I. L. and Lamata, L. and Solano, E.},
  journal = {Phys. Rev. Lett.},
  volume = {115},
  issue = {24},
  pages = {240502},
  numpages = {6},
  year = {2015},
  month = {Dec},
  publisher = {American Physical Society},
  doi = {10.1103/PhysRevLett.115.240502},
  url = {https://link.aps.org/doi/10.1103/PhysRevLett.115.240502}
}

@article{Mendicelli23,
    author = "Mendicelli, Emanuele",
    title = "{Investigating how to simulate gauge theories on a quantum computer}",
    eprint = "2308.15421",
    journal = "arXiv",
    primaryClass = "hep-lat",
    school = "York U., Canada",
    year = "2023"
}

@article{Lee2023EnergyCorrelators,
  title = {Quantum computing for energy correlators},
  author = {Lee, Kyle and Turro, Francesco and Yao, Xiaojun},
  journal = {Physical Review Letters},
  volume = {131},
  number = {17},
  pages = {171902},
  year = {2023},
  doi = {10.1103/PhysRevLett.131.171902},
  publisher = {American Physical Society}
}

@article{KlcoSavageStryker2020_SU2OneDimQuantum,
  title        = {SU(2) non‐Abelian gauge field theory in one dimension on digital quantum computers},
  author       = {Klco, Natalie and Savage, Martin J. and Stryker, Jesse R.},
  journal      = {Physical Review D},
  volume       = {101},
  number       = {7},
  pages        = {074512},
  year         = {2020},
  doi          = {10.1103/PhysRevD.101.074512},
  eprint       = {1908.06935},
  archivePrefix = {arXiv},
  primaryClass = {quant‑ph}
}

@article{Zache2023QuantumSpinNetwork,
  title = {Quantum and Classical Spin‑Network Algorithms for q‑Deformed Kogut‑Susskind Gauge Theories},
  author = {Zache, Torsten V. and Gonzalez‑Cuadra, Daniel and Zoller, Peter},
  journal = {Physical Review Letters},
  volume = {131},
  number = {17},
  pages = {171902},
  year = {2023},
  doi = {10.1103/PhysRevLett.131.171902},
  url = {https://doi.org/10.1103/PhysRevLett.131.171902}
}

@article{PeterWeyl27,
  author    = {F. Peter and H. Weyl},
  title     = {Die Vollständigkeit der primitiven Darstellungen einer geschlossenen kontinuierlichen Gruppe},
  journal   = {Mathematische Annalen},
  year      = {1927},
  volume    = {97},
  pages     = {737--755},
  doi       = {10.1007/BF01447866}
}

@article{Wigner59,
  author    = {Eugene P. Wigner},
  title     = {Group Theory and its Application to the Quantum Mechanics of Atomic Spectra},
  journal = {Academic Press},
  year      = {1959}
}

@book{BrinkSatchler1993,
  author    = {D. M. Brink and G. R. Satchler},
  title     = {Angular Momentum},
  edition   = {3rd},
  publisher = {Oxford University Press},
  year      = {1993}
}

@incollection{Penrose1971,
  author    = {Roger Penrose},
  title     = {Angular momentum: An approach to combinatorial space-time},
  booktitle = {Quantum Theory and Beyond},
  editor    = {T. Bastin},
  publisher = {Cambridge University Press},
  year      = {1971},
  pages     = {151--180}
}

@article{RovelliSmolin1995,
  author    = {Carlo Rovelli and Lee Smolin},
  title     = {Spin networks and quantum gravity},
  journal   = {Physical Review D},
  year      = {1995},
  volume    = {52},
  pages     = {5743--5759},
  doi       = {10.1103/PhysRevD.52.5743}
}

@article{Baez1996,
  author    = {John C. Baez},
  title     = {Spin network states in gauge theory},
  journal   = {Advances in Mathematics},
  year      = {1996},
  volume    = {117},
  pages     = {253--272},
  doi       = {10.1006/aima.1996.0012}
}

@article{Peruzzo2014,
  author    = {Alberto Peruzzo and Jarrod McClean and Peter Shadbolt and Man-Hong Yung and Xiao-Qi Zhou and Peter J. Love and Alán Aspuru-Guzik and Jeremy L. O'Brien},
  title     = {A variational eigenvalue solver on a photonic quantum processor},
  journal   = {Nature Communications},
  year      = {2014},
  volume    = {5},
  pages     = {4213},
  doi       = {10.1038/ncomms5213}
}

@article{Kandala2017,
  author       = {Kandala, Abhinav and Mezzacapo, Antonio and Temme, Kristan and Takita, Maika and Brink, Markus and Chow, Jerry M. and Gambetta, Jay M.},
  title        = {Hardware-efficient variational quantum eigensolver for small molecules and quantum magnets},
  journal      = {Nature},
  volume       = {549},
  number       = {7671},
  pages        = {242--246},
  year         = {2017},
  doi          = {10.1038/nature23879},
  url          = {https://doi.org/10.1038/nature23879},
  issn         = {1476-4687}
}

@article{McClean2016,
  author       = {McClean, Jarrod R. and Romero, Jonathan and Babbush, Ryan and Aspuru-Guzik, Alán},
  title        = {The theory of variational hybrid quantum-classical algorithms},
  journal      = {New Journal of Physics},
  volume       = {18},
  number       = {2},
  pages        = {023023},
  year         = {2016},
  doi          = {10.1088/1367-2630/18/2/023023},
  url          = {https://iopscience.iop.org/article/10.1088/1367-2630/18/2/023023}
}

@article{McClean2018,
  author    = {McClean, Jarrod R. and Boixo, Sergio and Smelyanskiy, Vadim N. and Babbush, Ryan and Neven, Hartmut},
  title     = {Barren plateaus in quantum neural network training landscapes},
  journal   = {Nature Communications},
  volume    = {9},
  number    = {1},
  pages     = {4812},
  year      = {2018},
  doi       = {10.1038/s41467-018-07090-4},
  url       = {https://www.nature.com/articles/s41467-018-07090-4}
}

@article{Holmes2022,
  author    = {Zoë Holmes and Kunal Sharma and Marco Cerezo and Patrick J. Coles},
  title     = {Connecting Ansatz Expressibility to Gradient Magnitudes and Barren Plateaus},
  journal   = {PRX Quantum},
  volume    = {3},
  number    = {1},
  pages     = {010313},
  year      = {2022},
  doi       = {10.1103/PRXQuantum.3.010313},
}

@article{Wang2021,
  author    = {Samson Wang and Enrico Fontana and M. Cerezo and Kunal Sharma and Akira Sone and Lukasz Cincio and Patrick J. Coles},
  title     = {Noise-induced barren plateaus in variational quantum algorithms},
  journal   = {Nature Communications},
  volume    = {12},
  number    = {1},
  pages     = {6961},
  year      = {2021},
  doi       = {10.1038/s41467-021-27045-6},
  url       = {https://www.nature.com/articles/s41467-021-27045-6}
}

@article{Cerezo2021,
  author    = {M. Cerezo and A. Sone and T. Volkoff and L. Cincio and P. J. Coles},
  title     = {Cost function dependent barren plateaus in shallow parametrized quantum circuits},
  journal   = {Nature Communications},
  volume    = {12},
  number    = {1},
  pages     = {1791},
  year      = {2021},
  doi       = {10.1038/s41467-021-21728-w},
  url       = {https://www.nature.com/articles/s41467-021-21728-w}
}

@article{Mathis2020,
   title={Toward scalable simulations of lattice gauge theories on quantum computers},
   volume={102},
   ISSN={2470-0029},
   url={http://dx.doi.org/10.1103/PhysRevD.102.094501},
   DOI={10.1103/physrevd.102.094501},
   number={9},
   journal={Physical Review D},
   publisher={American Physical Society (APS)},
   author={Mathis, Simon V. and Mazzola, Guglielmo and Tavernelli, Ivano},
   year={2020},
   month=nov }

@article{Wecker2015,
   title={Progress towards practical quantum variational algorithms},
   volume={92},
   ISSN={1094-1622},
   url={http://dx.doi.org/10.1103/PhysRevA.92.042303},
   DOI={10.1103/physreva.92.042303},
   number={4},
   journal={Physical Review A},
   publisher={American Physical Society (APS)},
   author={Wecker, Dave and Hastings, Matthew B. and Troyer, Matthias},
   year={2015},
   month=oct }

@misc{zhu2024,
      title={Optimizing Shot Assignment in Variational Quantum Eigensolver Measurement}, 
      author={Linghua Zhu and Senwei Liang and Chao Yang and Xiaosong Li},
      year={2024},
      eprint={2307.06504},
      archivePrefix={arXiv},
      primaryClass={quant-ph},
      url={https://arxiv.org/abs/2307.06504}, 
}

@article{Magnifico24,
  title        = {Tensor Networks for Lattice Gauge Theories beyond one dimension: a Roadmap},
  author       = {Giuseppe Magnifico and Giovanni Cataldi and Marco Rigobello and Peter Majcen and Daniel Jaschke and Pietro Silvi and Simone Montangero},
  journal      = {arXiv preprint arXiv:2407.03058},
  year         = {2024},
  eprint       = {2407.03058},
  archivePrefix= {arXiv},
  primaryClass = {hep‑lat}
}

@article{Cataldi23,
  title        = {(2+1)D SU(2) Yang--Mills Lattice Gauge Theory at finite density with tensor networks},
  author       = {Giovanni Cataldi and Giuseppe Magnifico and Pietro Silvi and Simone Montangero},
  journal      = {arXiv preprint arXiv:2307.09396},
  year         = {2023},
  eprint       = {2307.09396},
  archivePrefix= {arXiv},
  primaryClass = {hep‑lat}
}

@article{Marolf2000,
  author = {Donald Marolf},
  title = {Group averaging and refined algebraic quantization: Where are we now?},
  journal = {arXiv preprint gr-qc/0011112},
  year = {2000},
  archivePrefix = {arXiv},
  eprint = {gr-qc/0011112}
}

@article{KaminskiLiegener2020,
  title        = {Symmetry restriction and its application to gravity},
  author       = {Kami{\'n}ski, Wojciech and Liegener, Klaus},
  journal      = {arXiv preprint arXiv:2009.06311},
  year         = {2020},
  eprint       = {2009.06311},
  archivePrefix= {arXiv},
  primaryClass = {gr-qc},
  doi          = {10.1088/1361-6382/abdf29},
  journalref   = {Class. Quantum Grav. \textbf{38}, 065013 (2021)},
  note         = {Submitted on 14 September 2020; 69 pages}
}

@article{Ciavarella2025,
  title        = {Truncation uncertainties for accurate quantum simulations of lattice gauge theories},
  author       = {Ciavarella, Anthony N. and Hariprakash, Siddharth and Halimeh, Jad C. and Bauer, Christian W.},
  journal      = {arXiv preprint arXiv:2508.00061},
  year         = {2025},
  primaryClass = {quant‑ph},
  crossList    = {hep-lat, hep-ph, nucl-th},
  url          = {https://arxiv.org/abs/2508.00061},
  note         = {Submitted 31 Jul 2025; 27 pages, 8 figures}
}

@article{wigner1940,
  title     = {On the Matrices Which Reduce the Kronecker Products of Representations of Simply Reducible Groups},
  author    = {Wigner, E. P.},
  journal   = {Proceedings of the National Academy of Sciences of the USA},
  volume    = {26},
  number    = {2},
  pages     = {106--110},
  year      = {1940},
  doi       = {10.1073/pnas.26.2.106}
}

@article{racah1942,
  title     = {Theory of Complex Spectra. II},
  author    = {Racah, G.},
  journal   = {Physical Review},
  volume    = {62},
  number    = {9-10},
  pages     = {438--462},
  year      = {1942},
  doi       = {10.1103/PhysRev.62.438}
}

@article{Mattern2024,
  title     = {Simulating the 1-Vertex Model of SU(2) Lattice Gauge
Theory on Near Term Quantum Computers},
  author    = {Mattern, D.},
  journal   = {Bachelor thesis, TUM},
  volume    = {},
  number    = {},
  pages     = {},
  year      = {2024},
  doi       = {}
}

@article{Cochran2025,
  title        = {Visualizing Dynamics of Charges and Strings in (2+1)D Lattice Gauge Theories},
  author       = {Cochran, Tyler A. and Jobst, Bernhard and Rosenberg, Eliott and Lensky, Yuri D. and Gyawali, Gaurav and Eassa, Norhan and Will, Melissa and Abanin, Dmitry and Acharya, Rajeev and Aghababaie Beni, Laleh and Andersen, Trond I. and Ansmann, Markus and Arute, Frank and Arya, Kunal and Asfaw, Abraham and Atalaya, Juan and Babbush, Ryan and Ballard, Brian and ...},
  journal      = {Nature},
  volume       = {642},
  pages        = {315--320},
  year         = {2025},
  doi          = {10.1038/s41586-025-08999-9}
}

@article{Dusuel2015,
  title        = {Mean-field ansatz for topological phases with string tension},
  author       = {Dusuel, S. and Vidal, J.},
  journal      = {Physical Review B},
  volume       = {92},
  pages        = {125150},
  year         = {2015},
  doi          = {10.1103/PhysRevB.92.125150}
}

@article{OMalley2016UCCSD,
  title   = {Scalable quantum simulation of molecular energies},
  author  = {O’Malley, Peter J. J. and Babbush, Ryan and Kivlichan, Ian D. and Romero, Jonathan and McClean, Jarrod R. and Barends, Rami and Kelly, Julian and Roushan, Pedram and Tranter, Andrew and Ding, Nan and Campbell, Earl and Chen, Yu and Chen, Zijun and Chiaro, Ben and Dunsworth, Andrew and Jeffrey, Evan and Megrant, Anthony and Mutus, Josh and Neeley, Matthew and Neill, Charles and Quintana, Chris and Sank, Daniel and Vainsencher, Adam and Wenner, Justin and White, Thomas and Love, Peter J. and Aspuru-Guzik, Alán and Cleland, Andrew N. and Martinis, John M.},
  journal = {Physical Review X},
  volume  = {6},
  number  = {3},
  pages   = {031007},
  year    = {2016},
  doi     = {10.1103/PhysRevX.6.031007}
}

@misc{ToffoliDecomposition,
  title = {Elementary gates for quantum computation},
  author = {Barenco, Adriano and Bennett, Charles H. and Cleve, Richard and DiVincenzo, David P. and Margolus, Norman and Shor, Peter and Sleator, Tycho and Smolin, John A. and Weinfurter, Harald},
  journal = {Phys. Rev. A},
  volume = {52},
  issue = {5},
  pages = {3457--3467},
  numpages = {0},
  year = {1995},
  month = {Nov},
  publisher = {American Physical Society},
  doi = {10.1103/PhysRevA.52.3457},
  url = {https://link.aps.org/doi/10.1103/PhysRevA.52.3457}
}

@article{powell1964_powell,
  author    = {M. J. D. Powell},
  title     = {An efficient method for finding the minimum of a function of several variables without calculating derivatives},
  journal   = {The Computer Journal},
  volume    = {7},
  number    = {2},
  pages     = {155--162},
  year      = {1964},
  doi       = {10.1093/comjnl/7.2.155}
}

@article{virtanen2020scipy_powell,
  author    = {Pauli Virtanen and Ralf Gommers and Travis E. Oliphant and Matt Haberland and Tyler Reddy and David Cournapeau and Evgeni Burovski and Pearu Peterson and Warren Weckesser and Jonathan Bright and St{\'e}fan J. van der Walt and Matthew Brett and Joshua Wilson and K. Jarrod Millman and Nikolay Mayorov and Andrew R. J. Nelson and Eric Jones and Robert Kern and Eric Larson and C J Carey and {\.I}lhan Polat and Yu Feng and Eric W. Moore and Jake VanderPlas and Denis Laxalde and Josef Perktold and Robert Cimrman and Ian Henriksen and E. A. Quintero and Charles R Harris and Anne M. Archibald and Ant{\^o}nio H. Ribeiro and Fabian Pedregosa and Paul van Mulbregt and SciPy 1.0 Contributors},
  title     = {SciPy 1.0: Fundamental Algorithms for Scientific Computing in Python},
  journal   = {Nature Methods},
  volume    = {17},
  pages     = {261--272},
  year      = {2020},
  doi       = {10.1038/s41592-019-0686-2}
}

@article{Qiskit2023,
  author       = {Watkins, Alex and Sivak, Michael and Moreira, João and Capelluto, Louis and Cross, Andrew W. and Gambetta, Jay M. and McKay, David C. and Alexander, Thomas and et al.},
  title        = {Qiskit: An open-source framework for quantum computing},
  journal      = {Nature Reviews Physics},
  volume       = {5},
  pages        = {349--363},
  year         = {2023},
  doi          = {10.1038/s42254-023-00556-7},
  url          = {https://doi.org/10.1038/s42254-023-00556-7}
}

@article{huber2024parametric,
  title        = {Parametric multi‑element coupling architecture for coherent and dissipative control of superconducting qubits},
  author       = {Huber, G. B. P. and Roy, F. A. and Koch, L. and Tsitsilin, I. and Schirk, J. and Glaser, N. J. and Bruckmoser, N. and Schweizer, C. and Romeiro, J. and Krylov, G. and Singh, M. and Haslbeck, F. X. and Knudsen, M. and Marx, A. and Pfeiffer, F. and Schneider, C. and Wallner, F. and Bunch, D. and Richard, L. and S{\"o}dergren, L. and Liegener, K. and Werninghaus, M. and Filipp, S.},
  journal      = {arXiv preprint arXiv:2403.02203},
  year         = {2024},
  url          = {https://arxiv.org/abs/2403.02203},
  note         = {[quant-ph]},
}

@article{QiskitAer2020,
  title     = {Qiskit Aer: A high performance simulator framework for quantum circuits},
  author    = {Alexander, Thomas and Capelluto, Louis and Gacon, Julien and Granade, Christopher and Greenberg, Dinko and Guan, Weidong and Javadi-Abhari, Ali and Kanazawa, Naoki and Konovalov, Alexander and Liu, Douglas and Marques, Miguel and McKay, David C. and Moreira, Joao and Motta, Mario and Nation, Paul D. and Paz, Alejandro and Pistoia, Marco and Pritchett, Emily and Ray, William and Reed, Matt and Schwab-Molzen, Anna and Sertage, Ivan and Singh, Manish and Sivarajah, Seyon and Takita, Maika and Wood, Christopher J. and Wootton, James R. and Yu, Thomas and Gambetta, Jay M.},
  journal   = {arXiv preprint arXiv:2008.08571},
  year      = {2020},
  archivePrefix = {arXiv},
  eprint    = {2008.08571},
  primaryClass = {quant-ph},
  url       = {https://arxiv.org/abs/2008.08571}
}

@article{lang2017hamiltonianI,
  author    = {Thorsten Lang and Klaus Liegener and Thomas Thiemann},
  title     = {Hamiltonian Renormalisation I: Derivation from Osterwalder--Schrader Reconstruction},
  journal   = {arXiv preprint arXiv:1711.05685},
  year      = {2017},
  archivePrefix = {arXiv},
  eprint    = {1711.05685},
  primaryClass = {math-ph},
  url       = {https://arxiv.org/abs/1711.05685}
}

@article{Grimsley2019,
  author  = {Harper R. Grimsley and Sophia E. Economou and Edwin Barnes and Nicholas J. Mayhall},
  title   = {An adaptive variational algorithm for exact molecular simulations on a quantum computer},
  journal = {Nature Communications},
  volume  = {10},
  pages   = {3007},
  year    = {2019},
  doi     = {10.1038/s41467-019-10988-2},
  url     = {https://doi.org/10.1038/s41467-019-10988-2}
}

@misc{google2024willow,
  author       = {Google Quantum AI},
  title        = {Willow Spec Sheet},
  year         = {2024},
  url          = {https://quantumai.google/static/site-assets/downloads/willow-spec-sheet.pdf},
  note         = {Accessed: 2025-09-01}
}

@misc{postquantum_2025,
  author       = {Marin Ivezic},
  title        = {The Race Toward FTQC: Ocelot, Majorana, Willow, Heron, Zuchongzhi},
  year         = {2025},
  url          = {https://postquantum.com/quantum-computing/fault-tolerant-quantum-race/},
  note         = {Accessed: 2025-09-01}
}

@article{Sim2019expressibility,
  author = {Sukin Sim and Peter D. Johnson and Alán Aspuru-Guzik},
  title = {Expressibility and entangling capability of parameterized quantum circuits for hybrid quantum-classical algorithms},
  journal = {Advanced Quantum Technologies},
  volume = {2},
  number = {12},
  pages = {1900070},
  year = {2019},
  doi = {10.1002/qute.201900070},
  url = {https://doi.org/10.1002/qute.201900070}
}

@article{McArdle2018quantum,
  author = {Sam McArdle and Suguru Endo and Alán Aspuru-Guzik and Simon C. Benjamin and Xiao Yuan},
  title = {Quantum computational chemistry},
  journal = {Rev. Mod. Phys.},
  volume = {92},
  number = {1},
  pages = {015003},
  year = {2020},
  doi = {10.1103/RevModPhys.92.015003},
  url = {https://arxiv.org/abs/1808.10402},
  note = {arXiv:1808.10402 [quant-ph]}
}

@article{McArdle2019,
  author = {Sam McArdle and Xiao Yuan and Simon Benjamin},
  title = {Error-mitigated digital quantum simulation},
  journal = {Physical Review Letters},
  volume = {122},
  number = {18},
  pages = {180501},
  year = {2019},
  doi = {10.1103/PhysRevLett.122.180501},
  url = {https://arxiv.org/abs/1807.02467},
  note = {arXiv:1807.02467 [quant-ph]}
}

@article{BonetMonroig2018,
  author = {X. Bonet-Monroig and R. Sagastizabal and M. Singh and T. E. O'Brien},
  title = {Low-cost error mitigation by symmetry verification},
  journal = {Physical Review A},
  volume = {98},
  number = {6},
  pages = {062339},
  year = {2018},
  doi = {10.1103/PhysRevA.98.062339},
  url = {https://arxiv.org/abs/1807.10050},
  note = {arXiv:1807.10050 [quant-ph]}
}

@article{Ballini2025Symmetry,
  author       = {Edoardo, Ballini and Julius, Mildenberger and Matteo, M. Wauters and Philipp, Hauke},
  title        = {Symmetry Verification for Noisy Quantum Simulations of Non‑Abelian Lattice Gauge Theories},
  journal      = {arXiv},
  volume       = {abs/2412.07844},
  year         = {2025},
  month        = {Jul},
  archivePrefix= {arXiv},
  eprint       = {2412.07844},
  primaryClass = {quant-ph},
  note         = {Accepted in *Quantum* 2025-06-26; published under CC-BY 4.0}
}

@article{Bharti2022,
  author       = {Kishor Bharti and Alba Cervera-Lierta and Thi Ha Kyaw and Tobias Haug and Sumner Alperin-Lea and Abhinav Anand and Matthias Degroote and Hermanni Heimonen and Jakob S. Kottmann and Tim Menke and Wai-Keong Mok and Sukin Sim and Leong-Chuan Kwek and Alán Aspuru-Guzik},
  title        = {Noisy intermediate-scale quantum algorithms},
  journal      = {Reviews of Modern Physics},
  year         = {2022},
  volume       = {94},
  number       = {1},
  pages        = {015004},
  doi          = {10.1103/RevModPhys.94.015004},
  url          = {https://doi.org/10.1103/RevModPhys.94.015004}
}

@article{Temme2017,
  author       = {Kristan Temme and Sergey Bravyi and Jay M. Gambetta},
  title        = {Error Mitigation for Short-Depth Quantum Circuits},
  journal      = {Physical Review Letters},
  year         = {2017},
  volume       = {119},
  number       = {18},
  pages        = {180509},
  doi          = {10.1103/PhysRevLett.119.180509},
  url          = {https://doi.org/10.1103/PhysRevLett.119.180509}
}

@article{Calajo2024_SU2IonQudit,
  author       = {Giuseppe Calajó and Giuseppe Magnifico and Claire Edmunds and Martin Ringbauer and Simone Montangero and Pietro Silvi},
  title        = {Digital Quantum Simulation of a (1+1)D SU(2) Lattice Gauge Theory with Ion Qudits},
  journal      = {PRX Quantum},
  volume       = {5},
  number       = {4},
  pages        = {040309},
  year         = {2024},
  doi          = {10.1103/PRXQuantum.5.040309},
  url          = {https://doi.org/10.1103/PRXQuantum.5.040309}
}
\appendix
\glsresetall % Reset acronyms, so they will be written in full form again the first them they are used in the appendix

\section{Computations on the gauge-invariant Hilbert space}
\label{app:TheoryDetails}
This appendix outlines the details behind the toy model motivated in \cref{Sec5:ToyModel} when described in terms of spin-network functions. The gauge theory is defined by holonomies $h\in$ SU (2) and gauge covariant fluxes $P\in \mathfrak{su}(2)$ subject to the \textit{holonomy-flux algebra}:
\begin{align}
    [h(e),P_I(e')] = h(e)\tau_I\delta_{ee'} \\
    [P_I(e),P_J(e')] = \epsilon_{IJK} P_K(e)\delta_{ee'}
\end{align}
and all other commutators vanish identically. The $\epsilon_{IJK}$ are the structure functions of the corresponding Lie algebra and $\tau_I$ a suitable basis, i.e., $\tau_I\in\mathfrak{su}(2)$ for the Lie group SU(2) used throughout this manuscript.\footnote{A common choice is $\tau_I = -i \sigma_I/2$ with $\sigma_I$ being the Pauli matrices.} We recall from the main text that the kinematical Hilbert space at any 6-valent vertex in a $d=3$ lattice can be expressed in terms of ingoing edges $e^-_k\in\gamma$ and outgoing edges $e^+_k\in\gamma$ with $k=1,2,3$. Using the Peter\&Weyl theorem \cite{PeterWeyl27} for each edge, the full tensor product Hilbert space takes the form $\H = \otimes_{e}\H_e$ with  basis elements 
\begin{align}
    \Pi_{k=1,2,3} D^{j_k^+}_{m_k^+n_k^+}(h^+_k) D^{j_k^-}_{m_k^-n_k^-}(h^-_k) \; .
\end{align}
The toy model under consideration is based on periodic boundary conditions, which we enforce by restricting to the following case (including a suitable relabeling)
\begin{align}
    j_k^+=j^-_k = j_k,\quad n^+_k=m^-_k=\nu_k, \\
    m_k^+=m_k,\quad n_k^-=n_k
\end{align}
and a summation over the twice appearing index $\nu_k$. Under this restriction, the system simplifies as its wave function will depend only on the three combinations of the form $h^+_kh^-_k\in$ SU(2), i.e., three independent copies over SU(2). Further, the restriction to the gauge-invariant Hilbert space $\H_G$ can now be performed in the usual manner by contracting with a six-valent intertwiner.
For the computation in this paper, we pick a specific basis of intertwiners on the cubic lattice labeled by the auxiliary irreps $\pi^v_{k},k=1,2,3$ over the unit in SU(2). Those and the irreps $j_k^\pm$ are contracted in suitable triples with Wigner-3j-symbols \cite{Wigner59}, i.e., the 3-valent intertwiners:
\begin{align} \label{eq:intertwiner3D}
 %\iota_{m_{1}m_2m_3,n_1,n_2,n_3}^{v,\pi_1,\pi_2,\pi_3}:=\sum_{\rho_1,\rho_2,\rho_3} \left(\begin{array}{ccc}
  %    \pi_1 & \pi_2 & \pi_3 \\
  %    \rho_1 & \rho_2 & \rho_3
 %\end{array}\right) \times\\
 %&\quad\quad\quad\times\left(\begin{array}{ccc}
 %     j_1 & j_1' & \pi_1 \\
 %     m_1 & n_1 & \rho_1
 %\end{array}\right)\left(\begin{array}{ccc}
 %     j_2 & j_2' & \pi_2 \\
 %     m_2 & n_2 & \rho_2
 %\end{array}\right)\left(\begin{array}{ccc}
 %     j_3 & j_3' & \pi_3 \\
 %     m_3 & n_3 & \rho_3
 %\end{array}\right)\nonumber
&\iota_{m_{1}m_2m_3,n_1,n_2,n_3}^{v,\pi_+,\pi_o,\pi_-}:=\sum_{\rho_+,\rho_o,\rho_-} \left(\begin{array}{ccc}
      j_1^+ & j_2^+ & \pi_+ \\
      m_1 & m_2 & \rho_+
 \end{array}\right) \times\\
 &\quad\quad\quad\times\left(\begin{array}{ccc}
      \pi_+ & j_3^+ & \pi_o \\
      \rho_+ & m_3 & \rho_o
 \end{array}\right)\left(\begin{array}{ccc}
      \pi_o & j_3^- & \pi_- \\
      \rho_o & n_3 & \rho_-
 \end{array}\right)\left(\begin{array}{ccc}
      j_1^- & j_2^- & \pi_- \\
      n_1 & n_2 & \rho_-
 \end{array}\right)\nonumber\;,
\end{align}
where $j_k^+$ stands for the irrep of the outgoing edge at $v$ and $j_k^-$ for the irrep of the ingoing edge in direction $k$, respectively.
%Any 6-valent intertwiner can be written in this form \cite{BrinkSatchler1993}, however, not all combinations of $\vec{j},\vec{\pi}$ yield finite gauge-invariant functions.  
%\begin{align}
%    \left(\begin{array}{ccc}
%      j_1 & j_2 & j_3 \\
%      m_1 & m_2 & m_3
% \end{array}\right)
%\end{align}
By this, we obtain the full spin-network functions describing the kinematical Hilbert space of the theory, characterized by a total of six quantum numbers:
\begin{align}
    T_{j_1,j_2,j_3,\pi_+,\pi_o,\pi_-} (\vec{h}) =: \langle \vec{h} | j_1,j_2,j_3 ; \pi_+,\pi_o,\pi_-\rangle\;.
\end{align}
To study a SU(2) Yang-Mills theory in this setting, we need to define a suitable discretization of the Hamiltonian on this lattice, which we construct with maximal similarity to the Kogut-Susskind Hamiltonian \cite{KogutSusskind75}:
\begin{align}
    \label{eq:KS-Hamiltonian_detail}
    &H= g_E\sum_e H_E(e) - g_B \sum_\Box H_B(\Box),\\
    &H_E(e) = \delta^{IJ} P_I(e)P_J(e),\\
    &H_B(\Box) = tr(h(\Box)+h(\Box)^\dagger)\;.
\end{align}
The fact that the minimal plaquettes are spanned by only two quantum numbers (instead of four on larger lattices) motivates recalling the regularization procedure of the continuum expression, explicitly of the curvature of the connection $F(A):= dA + A\wedge A$. Following the non-Abelian Stokes theorem, which states that
\begin{align}\label{eq:6val_spin_network}
    F^I_{ab}e^a e^bF^J_{cd}e^c e^d \delta_{IJ}\approx (1-tr(Re(h(\Box_{ab}))))/\epsilon^4
\end{align}
for any closed plaquette $\Box$ of size $\epsilon^2$ which starts at $v$ and has tangents oriented along the coordinate directions $a,b$ spanned by the normal vectors $(e^a)_k=\delta^a_{k}$. Note that there is no requirement to choose whether the tangents point in positive or negative direction, and we are free to choose a manifestly positive orientation at $v$, i.e., choose our plaquette to be oriented leaving along $j^+_a$ and returning along $j^+_b$. Approximating $F_{ab}F^{ab}$ in such a way leads to the following regularization of the Hamiltonian for the single 6-valent vertex graph:
\begin{align}\label{eq:6val_reg_Hamiltonian}
    H = (1-\lambda^2) \sum_a P^I(e_a^+)P_I(e_a^+) - 2\lambda^2 \sum_{ab} tr(Re(h(e_a^+)h(e_b^+)^\dagger)\;.
\end{align}
Note that for the coupling constants we used, a total rescaling of the Hamiltonian does not change the qualitative physics but only leads to a rescaling, essentially reducing us to a single parameter that specifies the relative weight between $H_E$ and $H_B$. Further, we compactified the interval of said weight $[0,\infty]$ to the finite region $\lambda\in[0,1]$. 
Computing the matrix elements of \cref{eq:6val_reg_Hamiltonian} in the basis of \cref{eq:6val_spin_network} can now be done using the usual angular momentum recoupling theory, see \cite{BrinkSatchler1993} for extensive guides. The basic equations used here are the recoupling of Wigner D-functions
\begin{align}\label{eq:HolonomyCoupling}
D_{m_1 n_1}^{(j_1)}(h) &D_{m_2 n_2}^{(j_2)}(h) = \sum_{j_3 = |j_1 - j_2|}^{j_1+j_2} \sum_{m_3, n_3} d_{j_3}(-1)^{n_3-m_3} \\
&\times
\begin{pmatrix}
    j_1 & j_2 & j_3 \\
    m_1 & m_2 & m_3
\end{pmatrix}
\begin{pmatrix}
    j_1 & j_2 & j_3 \\
    n_1 & n_2 & n_3
\end{pmatrix}
D_{-m_3, -n_3}^{(j_3)}(h)\; ,
\nonumber
\end{align}
the orthogonality relations between 3j-symbols
\begin{align}
\sum_{j_3, m_3} d_{j_3}
\begin{pmatrix}
j_1 & j_2 & j_3\\
m_1 & m_2 & m_3
\end{pmatrix}
\begin{pmatrix}
j_1 & j_2 & j_3\\
m_1' & m_2' & m_3
\end{pmatrix} 
& = \delta_{m_1, m_1'} \delta_{m_2, m_2'}
\label{eq:3jOrthogonality1} \\
\sum_{m_1, m_2} d_{j_3}
\begin{pmatrix}
j_1 & j_2 & j_3\\
m_1 & m_2 & m_3
\end{pmatrix}
\begin{pmatrix}
j_1 & j_2 & j_3'\\
m_1 & m_2 & m_3'
\end{pmatrix} 
& = \delta_{j_3, j_3'} \delta_{m_3,m_3'}
\label{eq:3jOrthogonality2}
\end{align}
and the definition of the 6j-symbol \cite{wigner1940,racah1942}:
\begin{align}
\label{eq:6jDef}
\begin{Bmatrix}
    a & b & c \\
    d & e & f  
\end{Bmatrix}
= \sum_{\substack{\alpha, \beta, \gamma, \delta, \\ \epsilon, \phi}} &(-1)^{(d+e+f+\delta+\epsilon+\phi)}
\begin{pmatrix}
    a & b & c \\
    \alpha & \beta & \phi
\end{pmatrix}\times\\
&\times
\begin{pmatrix}
    a & e & f \\
    \alpha & \epsilon & -\phi
\end{pmatrix}
\begin{pmatrix}
    d & b & f \\
    -\delta & \beta & \phi
\end{pmatrix}
\begin{pmatrix}
    d & e & c \\
    \delta & -\epsilon & \gamma
\end{pmatrix}\; .\nonumber
\end{align}
Upon performing this computation of matrix elements of the Hamiltonian explicitly, it becomes apparent that \cite{Mattern2024}
\begin{align}
    \langle \vec{j}',\pi_+',\pi_o',\pi_-'| H | \vec{j},\pi_+,\pi_o,\pi_- \rangle \propto \delta_{\pi_o',\pi_o}\;.
\end{align}
In other words, the Hamiltonian decomposes into superselection sectors, as the label $\pi_o$ never gets changed during the action of the Hamiltonian and is thus a conserved quantity. It follows that the eigenstates of the Hamiltonian must be supported in each superselection sector separately. In this article, we focus therefore on the superselection sector $\pi_o=0$ as the ground state is found in this sector for $\lambda=0$, in which case the solution is analytically known to be $j_k=0$, hence enforcing $\pi_k=0$ as well. Being interested in how this state changes with $\lambda\to 1$ requires numerical analysis.

Focusing in what what follows on $\pi_o=0$ will, due to the 3j-symbols in $\iota^{v_o,\pi_+\pi_o,\pi_-}$, enforce that $\pi_\pm=j^\pm_s=j_3$. This reduces the full physical Hilbert space of the superselection sector $\pi_o=0$ to one characterized by three quantum numbers $j_1,j_2,j_3$, and hence makes it isomorphic to the Hilbert space over three edges that are contracted with two distinct 3j-symbols, graphically envisioned as a $\Theta$-graph. The spin-network functions forming the basis of this space are hence labeled as
\begin{align}
    T_{\vec{j}}(\vec{h}) =: \braket{\vec{h} | j_1,j_2,j_3} = \braket{\vec{h} | \vec{j} }\;.
\end{align}
Using \cref{eq:HolonomyCoupling}-\cref{eq:6jDef}, we compute the matrix elements in this basis, i.e., the electric part:
\begin{align}
    \bra{j_1', j_2', j_3'} H_{E}(e_a) &\ket{j_1, j_2,j_3} = j_a(j_a+1) \delta_{j_1,j_1'}\delta_{j_2,j_2'}\delta_{j_3,j_3'}
\end{align}
and the magnetic part, here for the plaquette $a=1,b=2$: \cite{Mattern2024}
\begin{align}\label{eq:matrix_elements_mag}
 \bra{j_1', j_2', j_3'} H_{\Box_{12}} &\ket{j_1, j_2,j_3} = \sqrt{d_{j_1}d_{j_1'}d_{j_2}d_{j_2'}} \times\\
 &\times
\begin{Bmatrix}
j_1 & j_2 & j_3\\
j_2' & j_1' & 1/2
\end{Bmatrix}^2 \delta_{j_1, j_1'\pm 1/2} \delta_{j_2, j_2'\pm 1/2} \delta_{j_3, j_3'}\nonumber\; . 
\end{align}
We note that, in general, any gauge-invariant observable can be written in such an explicit form, making the spin-network basis a suitable tool for numerical investigations. A physically interesting example would be the 2-point correlator: This is a relevant observable in high-energy physics to track the correlation between different spatial regions of the system, symbolically referred to as the vacuum expectation value of the field operators $\braket{\Omega | A_a(x) A_b(y) | \Omega}$. However, in LGT, one needs to introduce a suitable regularization of $A_a(x)$ on the lattice and then, moreover, ensure that the complete observable is gauge-invariant. For both steps, there is a custom procedure in the literature: first, $A_a(v_o)\approx (h(e_a)-1)/\epsilon$ and any observable $\O_\gamma$ on the lattice $\gamma$ might be projected onto its gauge invariant version via the procedure of group averaging:
\begin{align}
    \O_\gamma^G( \vec{h}) := (\prod_{v\in\gamma}\int_{SU(2)} d\mu_H(g_v)) \O_\gamma (\{g_{e(0)} h_e g^\dagger_{e(1)}\}_{e\in\gamma})\;.
\end{align}
However, on large lattices, concatenating both procedures for the 2-point correlator $\braket{\Omega, A_a(x) A_b(y) \Omega}$ actually annihilates the expression when $x$ and $y$ are spatially separated, because of the open endpoints of their edges. Yet, for the single-vertex graph, it is necessary that $x=y=v_o$, and even if $a\neq b$, the starting and end points of the edges in both directions agree. This renders a non-trivial expression and \textit{allows to define a finite 2-point correlator on the single vertex graph}. Explicitly one obtains: \cite{Mattern2024}
\begin{align}\label{eq:gauge_inv_2p_correl}
    (A^I_a(v_o)A^J_b(v_o))^G=\frac{1}{3}\delta_{IJ}(tr(h(e_a)h(e_b)^\dagger)-tr(h(e_a)h(e_b))
\end{align}
To compute the matrix elements of this operator on the 6-valent vertex, note that $h(e_b)^\dagger$ means traversing the edge $e_b$ in the opposite direction of $h(e_b)$. Hence, no matter the orientation chosen, one of the two terms in \cref{eq:gauge_inv_2p_correl} will require a two-edge one leaving in direction $j^+$, the other in direction $j^-$, hence changing the quantum label $\pi_o$. As discussed previously, the ground state lies in the sector $\pi_o=0$, thus annihilating any operator that changes $\pi_o$. It follows that only the plaquette in the  positive direction will contribute non-trivially to the 2-point correlator, and the matrix elements are found to be:
\begin{align}
    \langle \Omega|  (A^I_a(v_o)A^J_b(v_o)^G |\Omega\rangle &=  \frac{1}{2} \delta_{IJ} \langle \Omega| H(\Box_{ab})|\Omega\rangle . 
\end{align}
Note that this corresponds to a single plaquette as it appears in the magnetic part of the Hamiltonian \cref{eq:matrix_elements_mag} and 2-point correlator; however, it vanishes under different regularizations of the Hamiltonian and on larger lattices. In this manuscript, it nonetheless serves as a well-defined measure for correlations in the system in different directions, while showing that different field species $I=1,2,3$ are completely uncorrelated.

\begin{figure}
    \centering
    \includegraphics[width=0.93\linewidth]{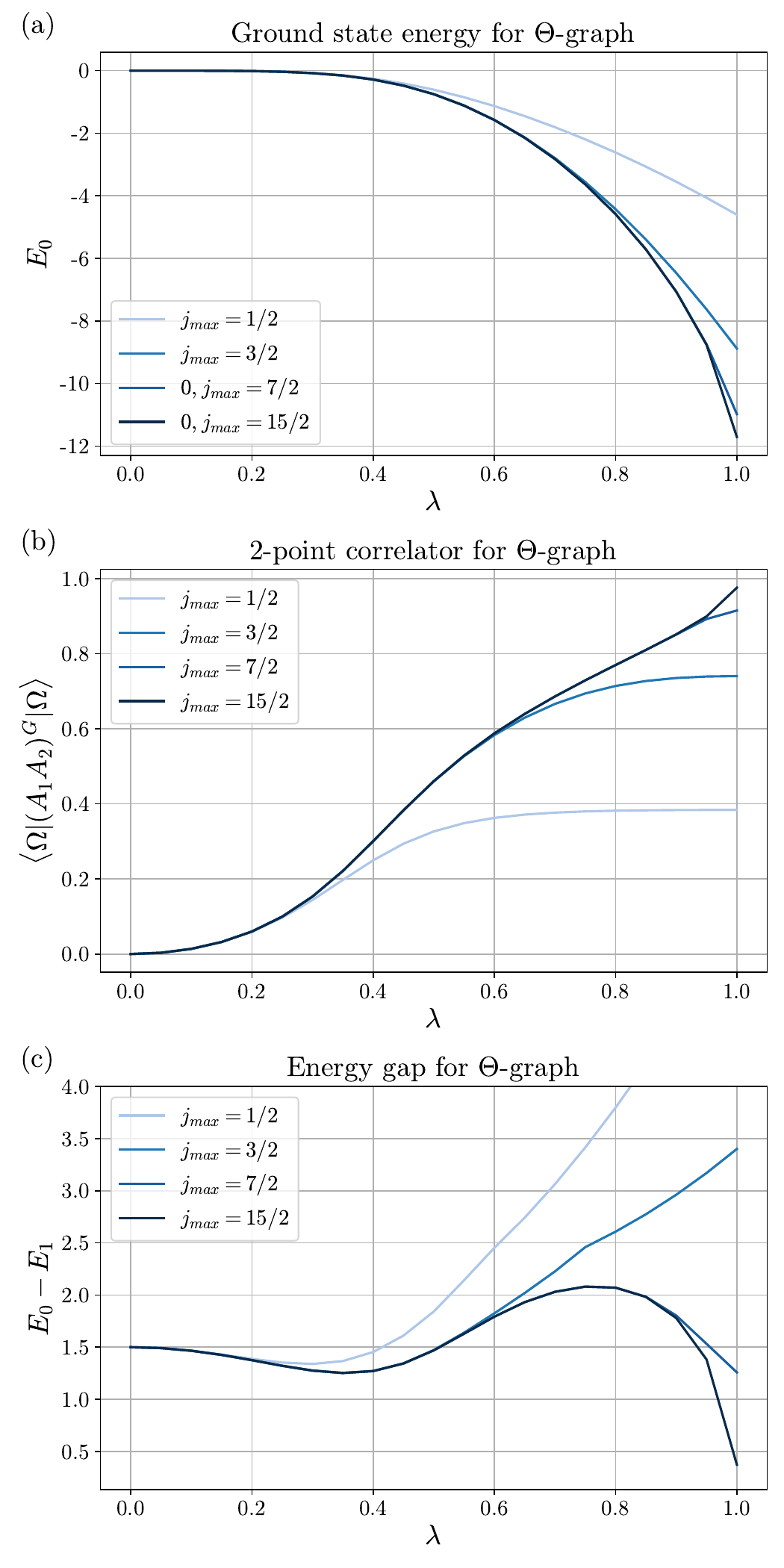}
    \caption{Numerical analysis of the cut-off dependency for the $\Theta$-graph ($\pi_o=0$ superselection sector of the single vertex). (a) Ground state energy varied with respect to the coupling parameter $\lambda$ converges around $j_{max}=4$. (b) 2-point correlation function indicates a phase transition around $\lambda\approx 0.5$ which is resolved with high accuracy at $j_{max}=3/2$. (c) The gap to the first excited state has a local minimum around the facsimile phase transition and vanishes for $\lambda\to1$ and $j_{max}\to \infty$.}
    \label{fig:ClassicalNumerics_CutOff}
\end{figure}

Having those expressions explicitly, one can perform a numerical analysis of the eigen system once a suitable cut-off $j_{max}$ has been introduced. The properties of the system will change when increasing $j_{max}$, therefore motivating its choice \textit{a posteriori} upon increasing it sufficiently until a certain convergence for the phenomena of interest has been achieved.\footnote{Currently, there is an ongoing effort in the field to determine the possible error due to finite cut-off a priori; however, these methods have not yet matured to apply to non-Abelian gauge theories \cite{Ciavarella2025}.
%https://arxiv.org/pdf/2508.00061
%https://ar5iv.labs.arxiv.org/html/1702.08838v1?utm_source=chatgpt.com
%https://arxiv.org/pdf/2506.16559
} Note that interest in different phenomena will require different cut-offs. In fact, analysis involving the ground state performs relatively mild with respect to the choice of $j_{max}$, as can be shown numerically: in \cref{fig:ClassicalNumerics_CutOff} (a) and (b), one can see that $j_{max}=3/2$ can capture the phase transition from uncorrelated to correlated phase (much more precise than $j_{max}=1/2$) motivating us in this paper to choose this cut-off, and use as proxy for the accuracy of the \gls{VQE} the difference at each $\lambda$ to the converged values when $j_{max}\to\infty$. Only in the strongly correlated regime does the cut-off $j_{max}=3/2$ fail to capture the behavior with high accuracy. This is expected because the analytic ground state of the system at $\lambda=1$ can be written formally as a Dirac-$\delta$-distribution over the group elements associated to each plaquette as
\begin{align}
\label{eq:weak_coupled_vacuum}
    \Omega_{\lambda=1}(\vec{h}_{\square}) = \prod_{\square} \delta(h_\square,e)
\end{align}
as the minimal energy corresponds to the sum of traces over each plaquette, which extremizes at $tr(e)=2$. For the $\Theta$-graph, this gives the vacuum energy of $E_{0,\lambda=1}=-12$ exactly, and we see that even at finite cut-off $j_{max}$ the energies rapidly converge towards this value.
However, not all features can be caught qualitatively well at these small cut-offs. E.g., we note that if one were interested in other physics such as the energy gap, a cut-off $j_{max}>2$ is mandatory to see the decrease in energy gap $\Delta=E_1-E_0$ in the highly correlated regime, which is neither captured by $j_{max}=1/2$ or $j_{max}=3/2$, see \cref{fig:ClassicalNumerics_CutOff} (c). In fact, higher cut-offs become relevant for the energy gap, as it is known to vanish for pure gauge theories for $\lambda=1$ because the trace over a plaquette takes on continuous values. Nonetheless, with the tools established in this paper, the regime of $\lambda<1$ is in principle accessible for \gls{VQE} methods on sufficiently large and accurate quantum devices using several qubits to store each irrep. We reserve this extension for future work.

\section{Systematic state preparation (SSP) circuits}
\label{app:Circuits}
In this appendix, we first introduce the general strategy to construct quantum circuits for systematic state preparation (SSP) of non-Abelian LGTs and, second, demonstrate this procedure in concrete examples for the lowest lying superselection sector of the 6-valent vertex studied in the main text and in \cite{Mattern2024}.

\subsection{General setup for SSP}
The SSP proposed in the main text is based on various developments in recent years to realize ground states of LGTs with shallow quantum circuits, while enriching it with the additional complexity encountered for non-Abelian LGTs. Notably, in contrast to theories such as $\mathbb{Z}_2$, the non-Abelian nature renders many methods, including the weight-adjustable loop ansatz (WALA) \cite{Cochran2025} and mean-field ansätze \cite{Dusuel2015}, unsuitable.\footnote{The added complexity comes from the bosonic nature of each link and the multi-fusion rules when coupling plaquettes together, see \cref{app:TheoryDetails}. The analytic ground state for $\lambda\to1$ is no longer the polarized product state, but rather a Dirac distribution peaked over the identity for each Wilson loop in the Hamiltonian, as seen in \cref{eq:weak_coupled_vacuum}.} 

\begin{figure}
    \centering
    \includegraphics[width=0.9\linewidth]{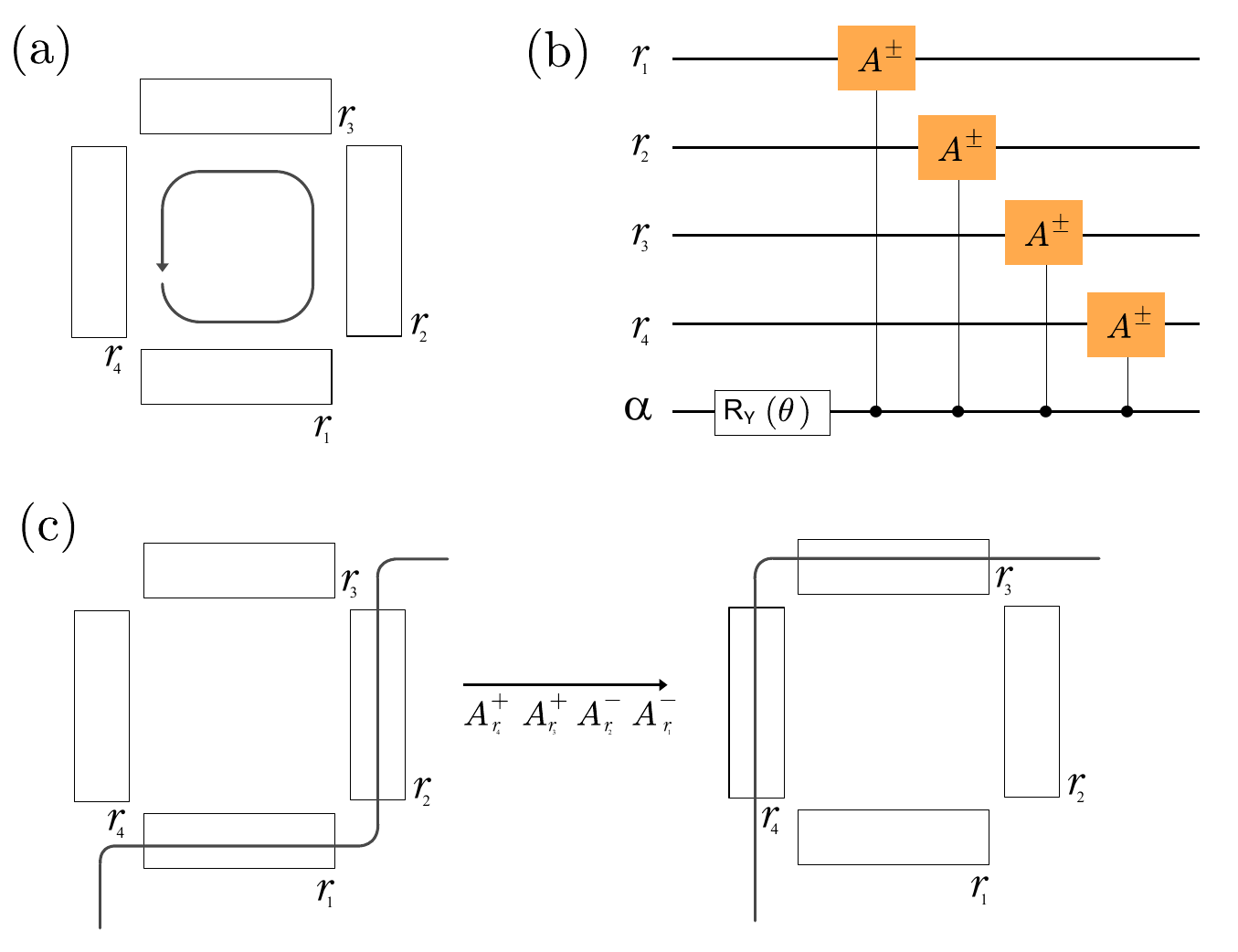}
    \caption{(a) Graphical representation of gauge-invariant excitations acting on qubit registers $r_k$ forming a minimal plaquette. According to the coupling rules for SU(2), the values in each register will be either raised or lowered. (b) Conditional action of the plaquette operator via a common ancilla qubit $\alpha$ and amplitude $\theta$. Any creation via $A^+$ conserves the gauge-invariant subspace at least until $j_{max}$ is reached. (c) Any other plaquette can be created via concatenating these operations, e.g., a flip of a string is achieved by $A^+_{r_4}A^+_{r_3}A^-_{r_2}A^-_{r_1}$.}
    \label{fig:Registers_Raising_Operations}
\end{figure}

We borrow, nevertheless, from WALA the idea of parsing conditional creations of full plaquettes into the state, thereby creating gauge-invariant excitations at all instances, i.e., elements of $\H_G$. This is beneficial against a dense \gls{HEA} \cite{Kandala2017} which samples the whole simulateable Hilbert space $\H_S$ on the quantum computer and suffers from a far larger number of required optimization parameters because in all so-far proposed non-Abelian simulations $\H_G \subset \H_S$. Said simulateable Hilbert space is defined by $\H_{S,n_q}:= \otimes_{n_q}\{0,1\}$ with $n_q$ many qubits and maps to the spin-network functions by assigning qubit registers of $r=\log_2 j_{max}$ qubits to each link $e$:
\begin{align}
    \H_e = \mathbb{Z}_{2j_{max}}:=\{0,...,2j_{max}-1\} \equiv \H_{S,r}\;.
\end{align}
Note that we apply the same procedure to each intertwiner label $\pi$. However, depending on the valency of the graph, the dimension of the intertwiners may change. In the basis chosen in \cref{eq:intertwiner3D}, we have $\pi_\pm\in\mathbb{Z}_{2j_{max}}$, but $\pi_o\in\mathbb{Z}_{3j_{max}}$.\footnote{A more symmetrical basis would be first coupling the edges in each direction together, i.e., $j_k^{\pm}$ to a common $\pi_k$, $k=1,2,3$, and then coupling all $\pi_k$ on a common 3j-symbol. Here it would be $\pi_k\in \mathbb{Z}_{2j_{max}}$, effectively reducing the required Hilbert space size. However, due to the presence of superselection sectors in the toy model, this manuscript works in the asymmetric intertwiner basis of \cref{eq:intertwiner3D}.} Now, parsing a gauge-invariant excitation, such as a closed curve, onto the quantum state requires a change in all the quantum registers belonging to said curve. Schematically, this is represented in \cref{fig:Registers_Raising_Operations}~(a) and requires an operation on each register, typically implemented with a common ancilla as in \cref{fig:Registers_Raising_Operations}~(b) to create a superposition of the gauge-invariant excitation being both applied and not applied.
In contrast to $\mathbb{Z}_2$, the operation to change the register is not a single CNOT gate, but requires raising/lowering the number represented on the register. I.e., we define the raising/lowering multi-qubit operations $A^\pm_r$ on an $r$-qubit register. Their explicit form depends on the ordering within the register, and we give a concrete example in the following subsection for $r=2$.
%Note that there is no need to use additional ancilla qubits, as, similar to the \gls{HEA}, also qubits within other registers may be used as control, as long as their excitation does not change non-trivially after the operation. 
The general idea is now to add successively on all minimal plaquettes excitations to the system, which, if concatenated, also distribute excitations among the lattice, as operations of the form $A^+_{r_4}A^+_{r_3}A^-_{r_2}A^-_{r_1}$ keep the total number of excitations constant, while flipping strings, see \cref{fig:Registers_Raising_Operations}~(c). Alternatively, suitable conditional SWAP operations along the lattice may be used to achieve a similar effect. Further, the addition of parametrized phase circuits in between them is necessary to allow for various signs on the states and incurs little additional 2-qubit gate cost. We mention in passing that there is a beneficial procedure to create the initial excitations, which benefit from the knowledge of several registers being in the ground state and therefore come at a reduced 2-qubit gate cost \cite{Mattern2024}. To parse this procedure onto large systems, an ordering of the circuit has to be chosen, similar to the procedure advocated in \cite{Cochran2025}. Yet, this ancilla-free procedure works only on the initial excitations. However, as non-Abelian LGT requires the option for arbitrary excitations on each link, the use of additional ancillas cannot be circumvented in general. Lastly, let us mention that the SSP procedure described here has a natural limit in terms of circuit depth, imposed by the cut-off $j_{max}$ in the system: to ensure unitarity of the operations $A^\pm_r$, it cannot always increase the register entry, and most implementations will be chosen such that $A^+ \ket{j_{max}}=\ket{0}$. One must therefore either avoid or pay additional attention to circuits that apply more than $j_{max}$-many raising operations on the same register, see the following subsection.

\subsection{Implementation of SSP on $\Theta$-graph}
To illustrate SSP, we note that for the $\Theta$-graph, the minimal plaquettes consist of two edges only, thereby simplifying the excitation creation in terms of actual 2-qubit gate count.
With the Hilbert space spanned by the states $\ket{j_1,j_2,j_3}$, which are labeled by the irrep over each of the three edges, we focus here on raising operations. E.g., on the pair 1,2, while leaving edge 3 invariant
\begin{align}
    j_1 \to j_1 +\frac{1}{2},\quad j_2 \to j_2 +\frac{1}{2},\quad j_3 \to j_3\;.
\end{align}
By restricting to $r=2$-qubit registers ($j_{max}=3/2$) for each edge, we denote the qubits constituting a register as follows: $j_k \equiv (j_{k,1},j_{k,2})$. Now, we can encode the raising operator as
\begin{align}
\vcenter{\includegraphics[scale=0.6]{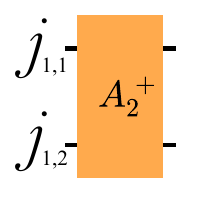}}\quad \hspace{-180pt}
= \begin{pmatrix}
0 & 0 & 0 & 1 \\
1 & 0 & 0 & 0 \\
0 & 1 & 0 & 0 \\
0 & 0 & 1 & 0
\end{pmatrix} \quad = \quad
\vcenter
{\includegraphics[scale=1]{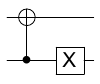}}
\end{align}
and $A^-_2:=(A^+_2)^\dagger$. To ensure unitarity, the state $\ket{11}$ must be mapped onto $\ket{00}$, which has the potential danger of mapping to a state violating gauge-invariance. One may either ensure in a given state preparation circuit that no $j$ is in the $\ket{3/2}$ state, or incorporate a check whether one of the irreps is in $\ket{3/2}$, i.e., the register being in the $\ket{11}$ state. This check can be accomplished by a Toffoli (CCNOT) gate, controlled by both qubits belonging to a $j$, which is used on an ancilla qubit. Then the excitation is only allowed if the ancilla qubits for both $j_1$ and $j_2$ are $\ket{0}$, which can be checked by using a CCRY gate towards a third ancilla qubit which serves as control for $A^+_{r_2}A^+_{r_1}$, i.e.:
\begin{align}
\includegraphics[scale=1]{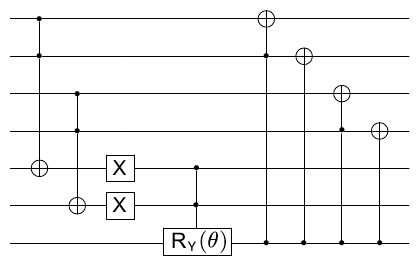}
\end{align}
%Adding an excitation this way is always allowed, but comes at an additional 2-qubit gate cost.
For this, all ancilla qubits must be in the $\ket{0}$ state initially. The first two ancilla qubits can be brought back to $\ket{0}$ afterward using two more Toffoli gates, but the third ancillary qubit can only be reset to $\ket{0}$ by measuring due to its entanglement with the qubit registers created by $CA^+_{r_2}A^+_{r_1}$. I.e., each such occurrence increases the likelihood of the SSP protocol failing and should therefore be used in moderation. A not-as-general, but cheaper way to add conditional excitations is when checking only the qubits encoding the half-integer digit of both registers. If both are in state $\ket{0}$, then rotating them into $\ket{1}$ always conserves the gauge-invariant subspace. This simplified excitation will be used for the conditional excitation in SSP4 below:
\begin{align}
\vcenter{\includegraphics[scale=1.05]{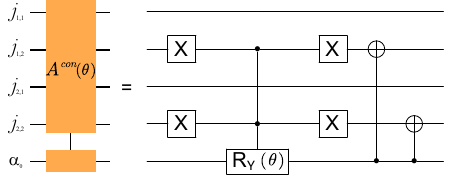}}
\end{align}
%\begin{figure}[h]
%    \centering
%    \includegraphics[scale=0.55]{figures/fig:ConditionalExcitationLower.png}
%    \caption{Circuit that adds an excitation to both $j_1$ and $j_2$ controlled by a parameter $\theta$ only if the lower qubits (tracking half-integer digits) are in state $\ket{0}$. This circuit is always allowed and a cheaper alternative to \cref{fig:ConditionalExcitation}.}
%    \label{fig:ConditionalExcitationLower}
%\end{figure}

Finally, we note that at the start of each SSP protocol, one knows that the qubits are in $\ket{0}$, which can be utilized to design an initial, cheaper creation of excitations, called in the following $A^{ini}$, without needing ancilla qubits and ignoring the higher entries of the qubit registers:
\begin{align}
\vcenter{\includegraphics[scale=0.7]{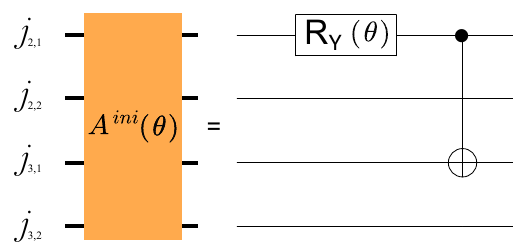}}
\end{align}

\begin{figure}[h]
    \centering
    \includegraphics[scale=1]{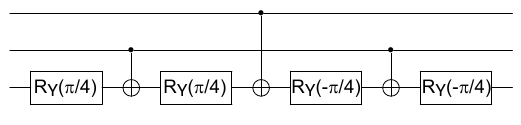}
    \caption{Cheap decomposition congruent to a Toffoli gate modulo phase shifts (i.e., a phase of $-1$ on the $\ket{101}$ state).}
    \label{fig:ToffoliDecompPhased}
\end{figure}

We briefly comment on the 2-qubit gate cost for $A^+_2$, which consists of 2 CNOT and 2 Toffoli gates. While the standard decomposition of a Toffoli gate uses 6 CNOT gates (bringing the total 2-qubit gate cost of this block to 14), there is a cheap modification congruent to the Toffoli gate modulo phase shifts \cite{ToffoliDecomposition}, in which the state $\ket{101}$ receives a relative phase of $-1$, see \cref{fig:ToffoliDecompPhased}. This decomposition uses only 3 CNOT gates, bringing the 2-qubit gate cost of a modified $\tilde A^+_2$ to 8. In our application, the additional phase is not problematic as it also preserves gauge invariance. In order to increase the fidelity in the presence of noise, we will use this phased decomposition in the following, unless mentioned otherwise.

Additionally, we mention that parameterized changes of the phase of states with respect to each other, which single-qubit gates can achieve, cannot violate gauge invariance and can always be added, due to the Hamiltonian being real and symmetric, its eigenstates will not carry a complex phase. Hence, all single-qubit rotations can be chosen to be parametrized by $R_Y(\theta)$ gates. Further, it is possible to conditionally move excitations around the lattice, thereby generating more symmetric states, which -- as outlined in the main text -- may prove helpful in generating translationally invariant states. This can be done with a sufficient concatenation of CSWAP gates such that gauge invariance in the final state is preserved. Again, this requirement simplifies in the present $\Theta$-graph, as the swap of just two irreps already presents a gauge-invariance-preserving operation, i.e., when $\ket{j_1 j_2 j_3}$ is a valid state, then $\ket{j_i j_j j_k}$ will also be a valid state for every permutation of $i,j,k \in \{1,2,3\}$. Doing such a conditional swap in a parameterized way is done by first rotating an ancilla qubit by an angle $\theta$ and then applying CSWAP gates controlled by the ancilla qubit onto the upper and lower qubit pair:
\begin{align}
    \vcenter{\includegraphics[scale=1.1]{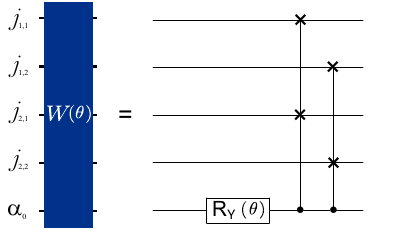}}
\end{align}
At the end of the circuit, each ancilla qubit is rotated by a Hadamard gate to create an entangled state, and post-selection ensures that the measured ancilla is in $\ket{0}$. Finally, we provide three explicit constructions based on the above-described rules. These three circuits, SSP2, SSP3, and SSP4, are used in the main text and pump up to $ 4, 6, 8$ excitations, respectively, into the system. At the cost of increasing the 2-qubit gate count and depending on more optimization parameters, higher accuracy can be achieved far into the highly correlated regime. In general, given a cut-off $j_{max}=3/2$, the upper maximum of gauge-invariant excitations on the $\Theta$-graph would be $8$ and hence not every state can be reached by the here expressed circuits. However, even for $\lambda=1$, we find that SSP4 captures the ground state with high accuracy. For small $\lambda$, mainly states with low $j$ values contribute, making the cheaper alternatives, SSP2 and SSP3, viable.

\section{Digital Twin}
\label{app:DigitalTwin}

\begin{figure}
    \centering
    \includegraphics[width=0.6\linewidth]{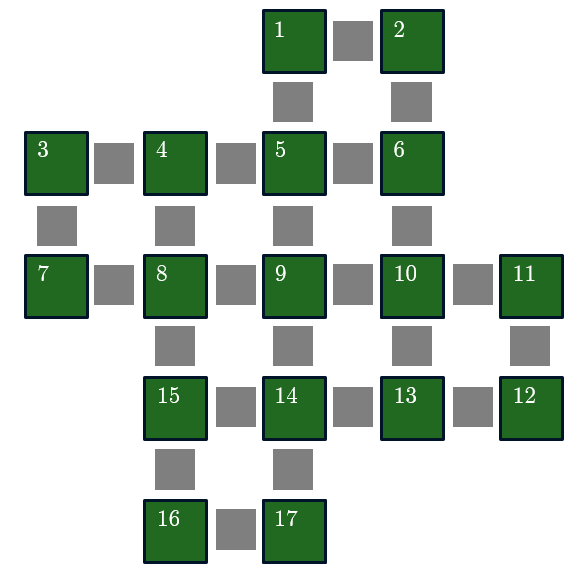}
    \caption{Layout of the chip with 17 qubits (green) and 24 couplers (grey) onto which the algorithms used in this manuscript have been transpiled for simulation.}
    \label{fig:17q_architecture}
\end{figure}

The simulations performed in \cref{Sec6:Implementation} have all been performed on an emulated qubit chip, inspired by contemporary superconducting qubit hardware. The underlying framework is the Qiskit Aer simulator \cite{Qiskit2023,QiskitAer2020} designed for modeling noisy quantum circuits. To incorporate the constraints originating from finite connectivity, we assume a square lattice topology to embed the up to 8 qubits for the SSP circuits. In particular, the 17-qubit and 24-coupler chip architecture presented in \cref{fig:17q_architecture} was used. The coupling map is passed to the built-in Qiskit transpiler at optimization level 3 to obtain an executable circuit that meets the minimum requirements with respect to swap operations. To make the noise model more in line with realistic hardware, we allowed only for $CZ$ gates as the native 2-qubit gate, which can be implemented on superconducting qubit hardware with tunable couplers \cite{huber2024parametric}. The characteristic noise profiles of all native gates were numerically simulated and passed as Kraus operators to the Qiskit Aer simulator, replacing the naive coherence-based noise maps.
In numerical simulations, it was found that the Kraus operator corresponding to the gate in \cite{huber2024parametric} features both systematic and stochastic errors, with systematic profiles dominating. This is, in fact, advantageous for the \gls{VQE} algorithm in the main text, as the optimization parameters in the circuits will be able to mitigate systematic shifts by adapting the optimization parameters. This built-in error mitigation is one of the reasons why applications based on \gls{VQE} hold promise, even for the current \gls{NISQ} era and the mid-sized quantum devices available at present. To be in agreement with the current achievable fidelities as those by \textit{IBM Quantum}, e.g.\@ Heron R2 \cite{postquantum_2025}, and \textit{Google Quantum}, e.g.\@ Willow \cite{google2024willow}, we fixed the values in the main text of this manuscript to 2-qubit gate error of $0.5\%$ and single qubit error of $0.05\%$.

\begin{figure}
    \centering
    \includegraphics[width=1\linewidth]{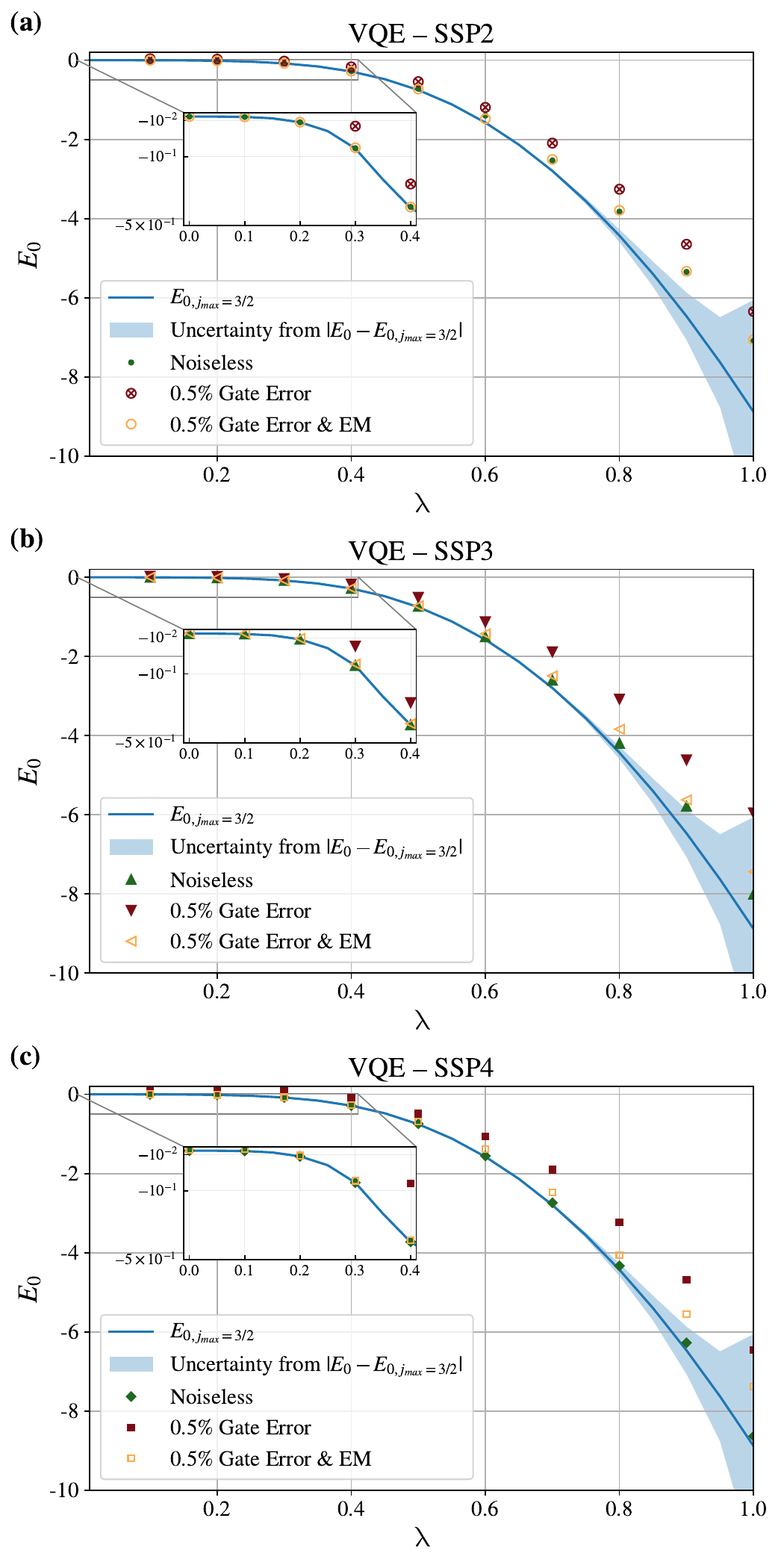}
    \caption{Best achieved minima for the energy expectation values for various SSP circuits. 
    The \gls{VQE} has been emulated in an ideal environment (green) and in the presence of noise from a 0.5\% 2-qubit gate error (red). Upon adding error mitigation (EM) protocols (orange), the accuracy comes close to the exact energies (blue) and falls within the uncertainty given by the finite cut-off $j_{max}=3/2$ (shaded region).}
    \label{fig:energy_exp_val}
\end{figure}

In \cref{fig:energy_exp_val}, we present exemplary results from \gls{VQE} runs on this digitial twin with different SSP variants.

\begin{widetext}
\begin{center}
\begin{figure}
    \includegraphics[width=\textwidth]{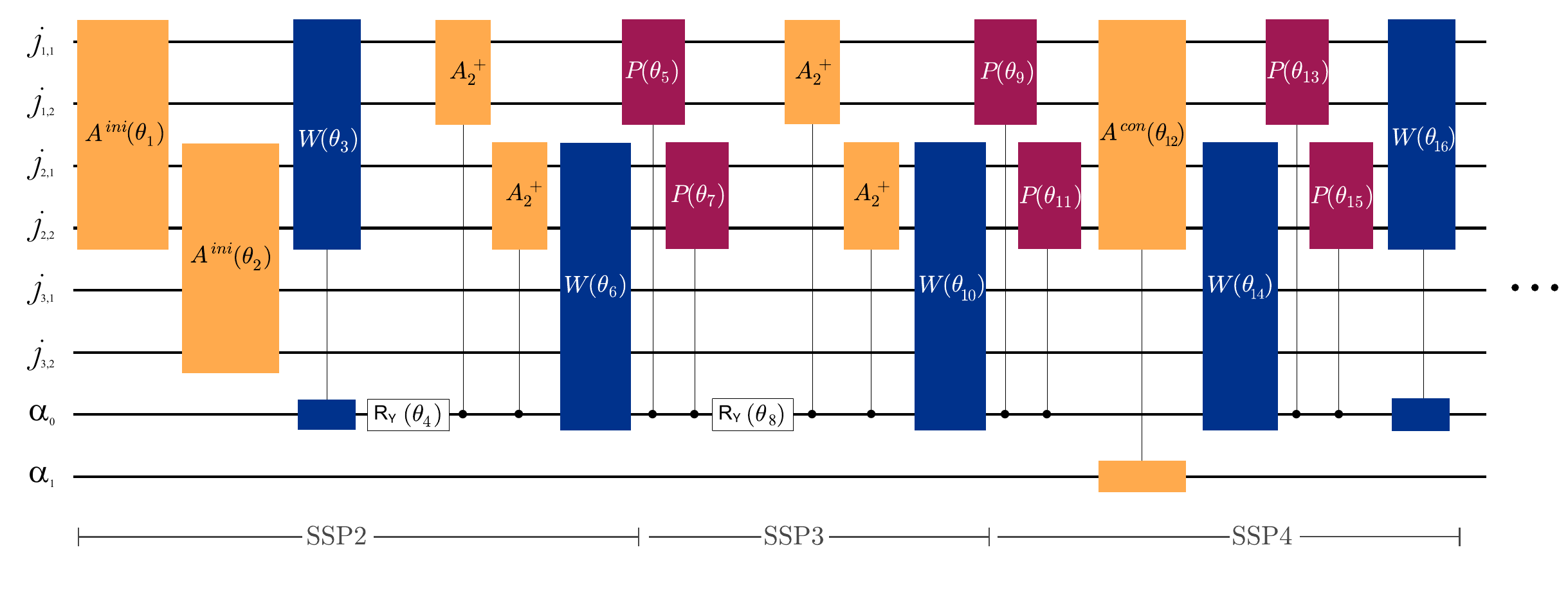}
    \caption{The full systematic state preparation protocol, which successively adds further excitations to the $\Theta$-graph. Cutting the circuit at various stages gives rise to the various SSP2, SSP3, and SSP4 used in the main text. Next to the excitation and swap operations defined in the main text, the circuit also features phase gates $P$, which are a concatenation of an $R_Y(\theta)$ and a CZ gate.
    }
    \label{fig:ThetagraphCircuits}
\end{figure}
\end{center}
\end{widetext}

\section{Comparison with HEA}
\label{app:HEA_VQE}
The SSP utilizes knowledge of the system in question by only creating gauge-invariant states and therefore scaling mildly in optimization parameter count. On the other hand, the \gls{HEA} relies only on a multitude of optimization parameters to sample a large part of the total Hilbert space and therefore achieves high expressibility while remaining agnostic to the system in question. The fact that \gls{HEA} is more dense and has a lower 2-qubit gate count than SSP raises the expectation that it may perform better in the \gls{NISQ} area, while -- once qubit errors become suppressed -- it is in danger of running into the barren plateau problem \cite{McClean2016,McClean2018,Wang2021,Cerezo2021}. In the following, we will investigate the impact of additional optimization parameters on the scaling in terms of additional cost-function evaluations associated with a vanishing gradient.

\begin{figure}
   \centering
    \includegraphics[width=0.93\linewidth]{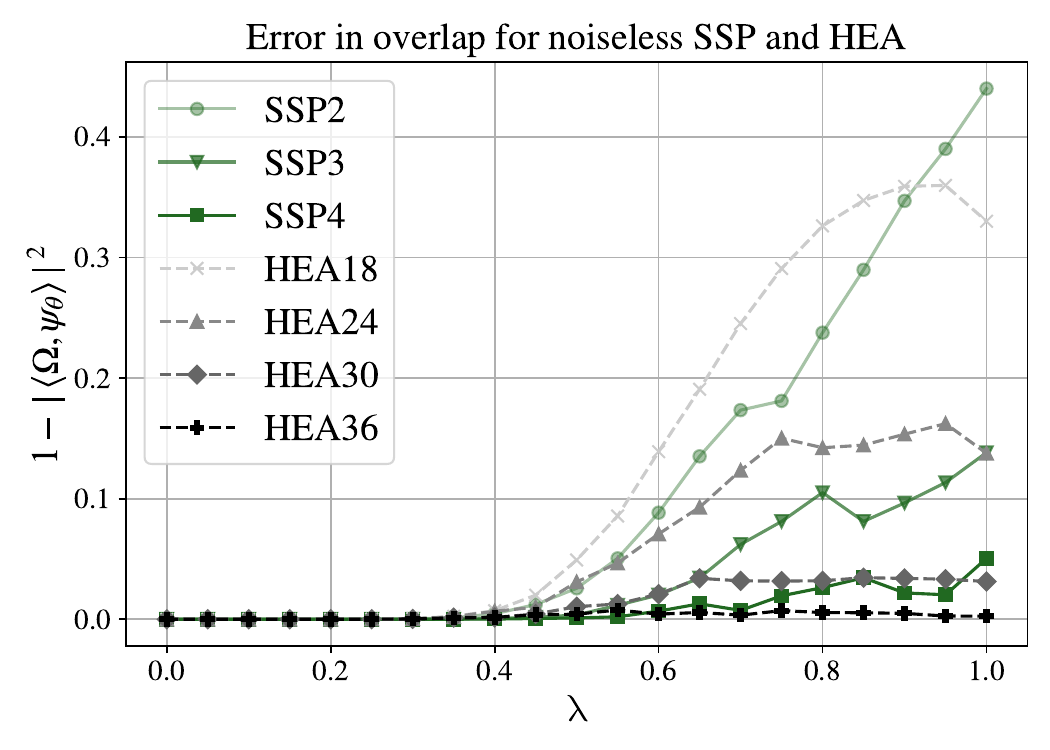}
    \caption{Comparison of different ans\"atze with respect to the optimal values reached in the ideal setting. The mismatch with the actual ground state $\Omega$ is captured in the error $1-|\langle \Omega, \psi_{\theta^*}\rangle|^2$. For ans\"atze reaching the same accuracy, SSP requires fewer optimization parameters than \gls{HEA}.
    }
    \label{fig:app_accurcacy_HEA_vs_SSP}
\end{figure}

\begin{figure}
   \centering
    \includegraphics[width=1\linewidth]{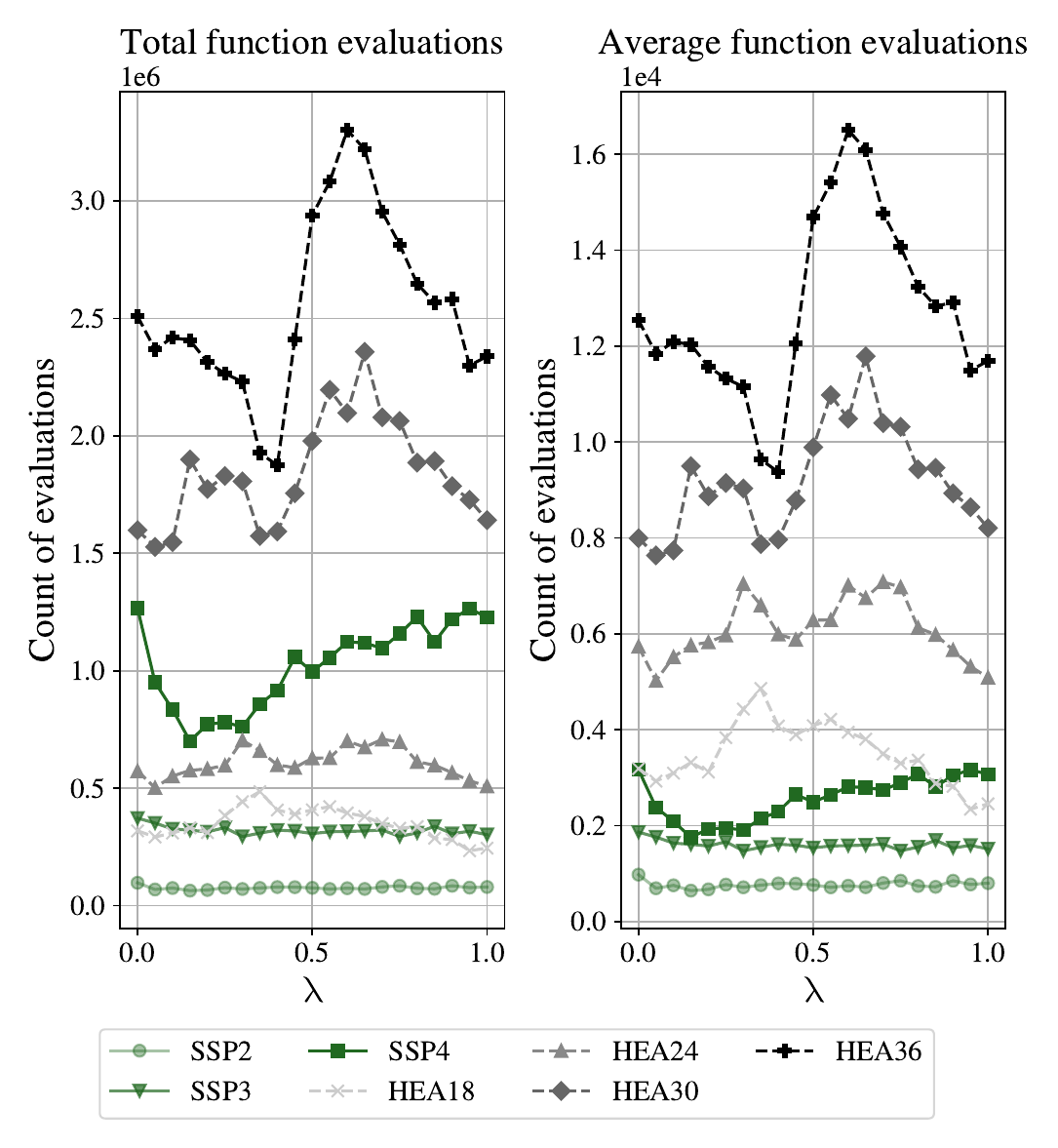}
    \caption{
    Number of required evaluations of the cost functions (i.e., measuring the expectation value for the Hamiltonian) for different ans\"atze. (Left) The total function evaluations count the number of candidates randomly scattered over the parameter landscape. The number of candidates is based on heuristic estimates and does not represent lower bounds. (Right) Average calls of the function evaluation for a single candidate to converge to its local minima.
    }
    \label{fig:app_func_evals_over_g}
\end{figure}

There are slight variants of \gls{HEA} used in the literature, see \cite{Sim2019expressibility} for a list of contemporary proposals. We focus in this paper on the architecture depicted in \cref{fig:AlgorithmAnsaetze}~(a), as relevant differences in their performance are not expected at the small system size considered in the present 6-valent vertex model. As both \gls{HEA} and SSP are families of algorithms parametrized by their depth, we will in our analysis compare different variants of SSP and \gls{HEA} \textit{a posteriori} depending on the accuracy they could reach in ideal (i.e., noise-free) simulations\footnote{This is the limit we are interested in, as the ongoing improvements in qubit gate fidelity spark the hope that quantum devices in the late \gls{NISQ} era will soon be able to deal with sufficiently deep algorithms when the barren plateau problem can become relevant.}. In \cref{fig:app_accurcacy_HEA_vs_SSP}, we show how accurate the algorithms are in preparing a state close to the exact ground state of the 6-valent vertex model at $j_{max}=3/2$ cut-off. This quantity is not obtainable in real simulations without an extensive quantum state tomography, but is easily obtainable on the digital twin emulator from \cref{app:DigitalTwin} utilized in this work. For small values of $\lambda$, all algorithms reach good approximations to the ground state, while in the regime of large $\lambda$, deeper algorithms perform better. Comparing different instances of \gls{HEA} and SSP in terms of their accuracy at $\lambda=1$ suggests primarily comparing the tuples (HEA18, SSP2), (HEA24, SSP3), and (HEA30, SSP4) as the elements in each of these sets perform similarly. We present a detailed comparison between the two families in \cref{table:KPI} with respect to their relevant metrics, such as qubit gate and optimization parameter counts. As expected, the number 2-qubit gates required for SSP is approximately a factor 5 larger than for \gls{HEA}, while the requirement for optimization parameters is approximately cut in half for the same accuracy level. For the optimizer, we considered two options: one method that utilizes the Hessian (BFGS) and another method based on gradients (Powell). BFGS uses fewer function evaluations than Powell, presumably due to the quicker convergence of using a Hessian. However, when noise is added to the simulation, BFGS does not reliably converge. This is assumed to be due to the Hessian being strongly impacted by faulty measurements. Powell was selected for our calculations regardless of noise presence for consistency. To get a rough idea how the increased optimization parameter count will be reflected in experiments, we study the convergence behaviour of the algorithms in terms of cost-function evaluations: In \cref{fig:app_func_evals_over_g} we show how long it took for algorithms with a given candidate count of $N=200$ until all candidates reached their local minimum, irrespective of how precise the result was. One can see that the number of required function evaluations increases non-linearly with parameter count. This feature is especially pronounced at $\lambda\approx 0.4$ where the fac-simile phase-transition of the system happens. Due to this scaling of needed function evaluations, SSP requires far fewer function evaluations than \gls{HEA} to reach a similar accuracy.
%, as can also be seen in \cref{fig:app_econvergence_over_func_evals}, where the convergence towards the global minimum over function evaluations (i.e., time equivalent) is plotted.
The increase in function evaluations hints at an eventual vanishing of gradients and indicates that the wall-clock time of \gls{VQE} with \gls{HEA} in LGT will scale unfavorably, encountering the barren plateau problem, which motivates the use of SSP in future applications.

% Removed for time being. Wait for finding of Alex

%\begin{figure}
%   \centering
%    \includegraphics[width=0.9\linewidth]{figures/fig_app_convergence.png}
%    \caption{\KL{Add convergence  over iteration plot for SSP and HEA at x=0.4 or at x=1?}}
%    \label{fig:app_econvergence_over_func_evals}
%\end{figure}

\section{Error Mitigation by Symmetry Verification}
\label{app:SymmVerif}

In this appendix, we discuss the methods of error mitigation (EM) used for the simulation of the \gls{VQE} algorithm on the digital twin in order to improve the noisy qubit gate operations. There is already an extensive body of literature on EM schemes; see \cite{McArdle2018quantum} for modern implementations. As we only added emulated noise to the system here, we refrain from applying all known error-mitigation schemes to the algorithm and keep those for later implementation.\footnote{Having these tools (such as, e.g., zero-noise-extrapolation, operator density renormalization \cite{McArdle2018quantum}) in backhand will prove valuable to raise qubit performance on actual hardware.} We are focusing on EM schemes that are either post-processing (PP), not as commonly employed or at least need additional input to work for non-Abelian lattice gauge theories.

\begin{figure}
    \centering
    \includegraphics[width=1\linewidth]{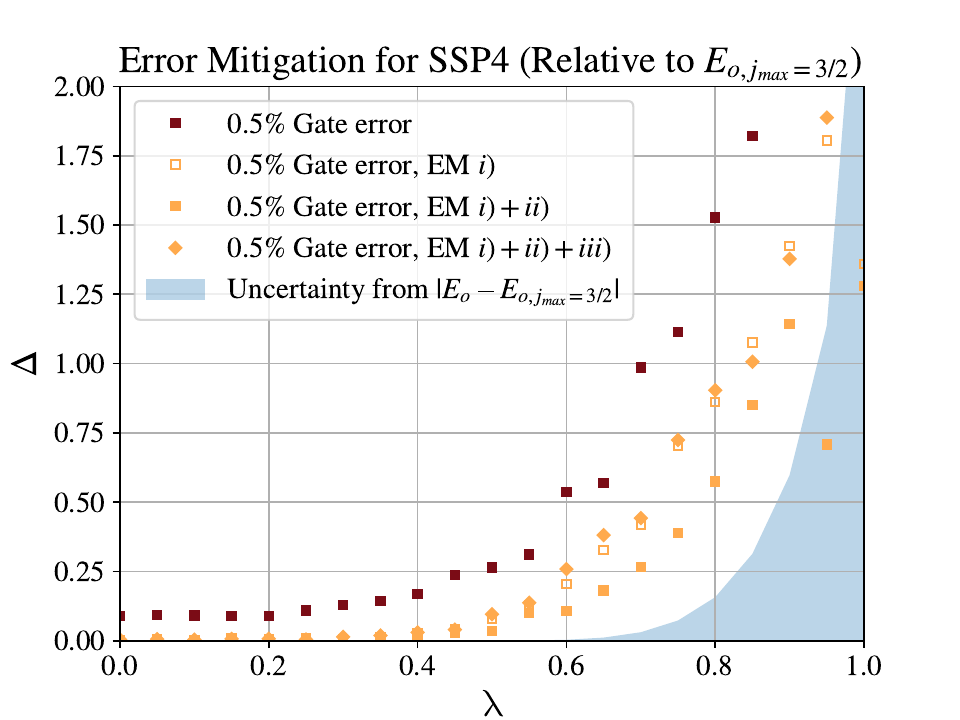}
    \caption{Effectiveness of various EM schemes exemplified on the error $\Delta$ for SSP4. Untreated noisy emulated data (2-qubit gate error $e=0.005$), shown in red, demonstrate an increase of the error for larger $\lambda$ values. Applying various EM methods allows for decreasing the error by obtaining better minima. Lastly, to reduce the spread in the results, we remove obvious outliers. This combination of methods $i)-iii)$ is used in the main text. Solid points indicate the median, and error bars indicate the standard deviation centered around the mean. In the main text of the manuscript, the used error-mitigation methods are $i)+ii)$.}
    \label{fig:app_EM}
\end{figure}

To see the impact of the EM schemes that we use, we show in \cref{fig:app_EM} how various combinations affect the results of emulating VQE on the 6-valent vertex model with the SSP4 algorithm emulated on the digital twin from \cref{app:DigitalTwin}. We show the derivation $\Delta$ from the expectation value for parameters $\theta^*=\theta^*(EM)$ obtained with various EM methods to exact ground state energy $E_{0,j_{max}=3/2}$ computed at finite cut-off $j_{max}=3/2$, i.e.\footnote{Note, that the Hamiltonian in the spin-network basis projects unphysical states to zero, hence yielding potentially a contribution to \cref{app:eq:error}. In general, adding a penalty term of sufficient weight on these terms can counteract the impact of any unphysical terms created by noise. In the toy model investigated here, the energy expectation values of the unphysical states in the electric part have been sufficiently shifted upwards.}
\begin{align}
\label{app:eq:error}
    \Delta = \langle\psi_{\theta^*}| H|\psi_{\theta^*}\rangle - E_{0,j_{max}=3/2}
\end{align}
To account for the spread of the result over many samples after optimization, we repeat the readout for $N=1.000$ samples and plot as solid points the median of the distribution and use error bars to indicate the standard deviation centered around the mean. The following EM methods have been investigated:\\

\begin{figure}
    \centering
    \includegraphics[width=0.45\linewidth]{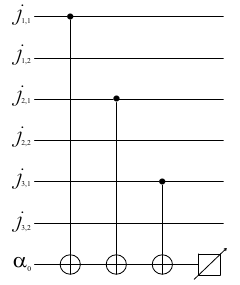}
    \caption{EM circuit used to perform in-bulk symmetry verification by enforcing condition \cref{eq:3j_1_2_condition}.
    }
    \label{fig:app-EM_circ}
\end{figure}

%N=400 i)
%N=401 i+ii)
%N=402 i+ii+iii)
%N=403 i+iii)

\textit{o) Complex phase rotation.} The ideal circuit profited from knowledge that, in pure SU(2) Yang-Mills theory, the Hamiltonian is real and symmetric, therefore no complex phases appear in the eigenstate, and the state preparation protocol benefited from the possibility to choose always only $R_Y(\theta)$ gates. However, in a noisy environment, dephasing occurs, introducing unwanted complex phases into the prepared state, which cannot be mitigated by parametrized $R_Y$ gates. To only minimally increase the optimization parameters, we will -- instead of adding to every $R_Y$ gate an accompanying $R_Z$ gate -- only minimally account for the dephasing and set at the end of the circuit the final single qubit rotations to $R_Z(\theta)$ gates. These parameters help mitigate any systematics from dephasing errors in the noise, while in the ideal setting, these gates will always insert phases of only $\pm1$. For this reason, we incorporate these complex phases into each circuit and have this EM technique automatically included in all ideal and noisy emulations.

\textit{i) PP symmetry verification for gauge-invariance.} In the seminal paper on symmetry verification for EM \cite{BonetMonroig2018,McArdle2019}, it was advocated to use the presence of symmetries of the system for both in-bulk and PP protocols.\footnote{That these tools translate straightforwardly to non-Abelian LGT was already observed in \cite{Ballini2025Symmetry}.} We focus first on the latter and note that gauge-invariance also presents a symmetry which is easy to verify by measuring a bit-string and asking whether it obeys the conditions \cref{eq:3j_1_2_condition,eq:3j_triangle_condition}. By performing this check at the end of the circuit, it comes for free without introducing additional noise; however, it is limited in its capabilities to mitigate multiple errors during state preparation or errors that map within the gauge-invariant subspace.

\textit{ii) PP rotation symmetry projector.} In general, any symmetry of a system may be used as a projection operator in PP. Given that the vacuum in LGT has translational symmetries (which have already been reduced in our toy model) and rotational symmetries by 90 degrees, the latter present a further possibility for applying symmetry verification. Yet, one needs to use caution, as already emphasized in \cite{McArdle2019}, because if a state preparation ansatz breaks the symmetry in intermediate steps, it is prone to introducing biases and needs to be carefully checked. For SSP, a rotation around axis $3$ will effectively switch $j_1\leftrightarrow j_2$ in the ground state; however, the incremental generation of excitations breaks this symmetry. However, due to the symmetrization of the state added during each SSP, the symmetry verification tool can still be used if one confirms \textit{a posteriori} that the cost function gets further minimized by applying this protocol. In our case, we do PP with respect to the following project, symmetrizing $j_1\leftrightarrow j_2$:
\begin{align*}
    P_{12}= (1 + S_{12})/2,\quad S_{12}=\frac{1}{4}\sum_{P,Q\in\{I,X,Y,Z\}}P_1Q_2P_3Q_4I_5I_6
    \;,
\end{align*}
with $S_{12}$ the SWAP of qubit registers 1 and 2, while leaving register 3 invariant.
The price to pay is that the obtained ground state has a bias towards this symmetry. Hence, expectation values of operators not obeying this symmetry will, in general, be faulty. For the application of this paper, we are interested in the 2-point correlator \cref{eq:gauge_inv_2p_correl}, which exists for any pair $a,b$ and is on the vacuum independent of this choice. Hence, we can use $P_{1\leftrightarrow2}$ if we restrict our analysis to $\langle \Omega| (A^I_1 A^J_2)^G|\Omega\rangle$, which has been done in the main text to produce \cref{fig:correlator}.

% Iii. In-bulk Symmetry verification
\textit{iii) In-bulk symmetry verification.}  Lastly, we mention a part of the standard tool-set for symmetry verification \cite{BonetMonroig2018,McArdle2019}, namely in-bulk symmetry verification measures. Here, one checks that a certain symmetry in the system is still preserved during the algorithm -- often by measurement -- and throws away runs where the measurement violates the condition. This procedure must be used in moderation, as successive applications will reduce the number of valid measurements and thus increase the wall-clock time for any algorithm. We use a check of the remaining $3j$ conditions to identify single qubit errors that map out of the gauge-invariant Hilbert space. Explicitly, equation \cref{eq:3j_1_2_condition} can be verified by checking the one qubit in each register that labels the half-integer value. A successive application of CNOT gates onto an ancilla qubit, as depicted in \cref{fig:app-EM_circ}, achieves that, and a run remains valid if the ancilla is measured in $|1\rangle$. However, we mark that this procedure introduces additional 2-qubit gates and therefore a potential further error source. It requires, therefore, a sufficiently good error rate to improve the overall error instead of worsening it. For the contemporary error sources emulated in this article, we simulated in \cref{fig:app_EM} the insertion of a  single in-bulk symmetry verification check in the middle of algorithm SSP4 (i.e., at the depth of the circuit corresponding roughly to the end of SSP3, compare \cref{fig:ThetagraphCircuits}). One can observe that, for the assumed error, the mitigation does not improve the measurement results. We thus refrain from using this method at the present stage for the error mitigation done in the main text.

% iV. Outlier removal
%\textit{iv) Outlier removal.}
%
%Throw away data that are far away (z-score threshold or quantile-based (IQR)
%These are classical data analysis techniques, and quantum papers may describe the behavior (e.g., “we discarded the top 5\% of deviating measurements”) without naming it as “Tukey’s method.”
%The methods are used under:
%“outlier rejection”
%“robust estimation.”
%“tail truncation.”
%“filtering anomalous samples.”
%
%\KL{Outlier removal in expectation value?}

\begin{widetext}

\begin{table}[h]
    \centering
    \setlength{\tabcolsep}{4pt} % optional: adjust column spacing
    \renewcommand{\arraystretch}{1.2} % optional: adjust row spacing
    \begin{tabular}{@{}l|cccccccc@{}} % 'l|' to separate first column
        Circuit & \# Ancillas &\# 1-Q Gates & \# 2-Q Gates & \# Layer &  \# Opt. Param. & \# Cand. & \# Avg. Eval. & Error (ideal) \\
        \toprule
        SSP2 & 1& 216 & 34 & 90 &  6  & 100 & 793 & $<$ 0.45  \\
        SSP3 & 1& 394 & 73 & 185 &  10  & 200 & 1,608 & $<$ 0.14  \\
        SSP4 & 2& 524 & 104 & 254 & 16 & 400 & 2,289 & $<$ 0.06 \\
        HEA18  & 0 & 104 & 10 & 24 & 18 & 100 & 4,074 & $<$ 0.34  \\
        HEA24  & 0 & 142 & 15 & 33 & 24 & 100 & 5,989 & $<$ 0.14  \\
        HEA30  &  0& 180 & 20 & 42 & 30 & 200 & 7,963 & $<$0.04  \\
        HEA36  & 0& 218 & 25 & 51 & 36 & 200 & 9,371 & $<$ 0.003  \\
        HEA42  & 0& 256 & 30 & 60 & 42 & 400 & 12,423 & $<$ 0.001  \\
    \end{tabular}
    \caption{Comparison of requirements for various circuit architectures and their projected (ideal) performance for the single vertex graph, which requires $6$ data qubits to store all information for $j_{max}=3/2$. The systematic state preparation (SSP) circuits are found to require higher qubit-gate count and more layers than circuits using the hardware efficient ansatz (HEA); however, they demonstrate a far tamer scaling in terms of optimization parameters, candidates, and evaluations of the cost function to reach similar ideal accuracy. This translates to a fast convergence of the optimizer and, therefore, alleviates the barren plateau problem. Comparisons of their performance are shown in \cref{fig:app_accurcacy_HEA_vs_SSP} and \cref{fig:app_func_evals_over_g}.}
    \label{table:KPI}
\end{table}
\end{widetext}

\end{document}